\documentclass{iopjournal}

\usepackage[version=4]{mhchem}
\usepackage{amssymb}
\usepackage{amsthm}
\usepackage{graphicx}
\usepackage{siunitx}
\usepackage{textcomp}
\usepackage[utf8]{inputenc}
\usepackage[T1]{fontenc}
\usepackage{mathptmx}
\usepackage{xcolor}
\usepackage{gensymb}
\usepackage[english]{babel}
\usepackage{url}
\usepackage{tabularx}
\usepackage{threeparttable}
\usepackage{booktabs}

\usepackage{multirow}
\usepackage{rotating}
\usepackage{makecell}
\usepackage{amsmath}
\usepackage[markup=underlined]{changes}
\usepackage[numbers,sort&compress]{natbib}
\usepackage{mathtools}
\usepackage{tensor}

\begin{document}

\articletype{Topical Review} 

\title{Hydrogen and Lithium Isotope Analysis via Optical Spectroscopy of Laser-Produced Plasmas}

\author{Elizabeth J. Kautz $^{1, 2,*}$\orcid{0000-0002-6338-9223}, Caleb Clark $^{1}$\orcid{0000-0002-7823-7214}, Nusrat Karim $^1$\orcid{0009-0004-3456-3853}, Mark C. Phillips$^{3,*}$\orcid{0000-0001-6566-7808}, and Sivanandan S. Harilal $^{2,*}$\orcid{0000-0003-2266-7976}}

\affil{$^1$North Carolina State University, Nuclear Engineering Department, Raleigh, North Carolina 27695, USA}

\affil{$^2$Pacific Northwest National Laboratory, Richland, Washington 99352, USA}

\affil{$^3$University of Arizona, Wyant College of Optical Sciences, Tucson, Arizona 85721, USA}

\affil{$^*$Authors to whom any correspondence should be addressed.}

\email{ekautz@ncsu.edu, mphillips@optics.arizona.edu, hari@pnnl.gov}


\begin{abstract} 
Accurate detection and quantification of light-element isotopes, particularly hydrogen (H) and lithium (Li), are essential across chemistry, materials science, geology, nuclear energy, forensics, and defense applications. Their isotopic compositions provide insight into fuel cycles, materials degradation, mass transport, and environmental processes. Hydrogen and Li isotopes are especially relevant in nuclear environments because neutron-induced reactions directly connect their isotopic inventories. Despite their importance, analysis of these light isotopes remains challenging due to their low atomic mass, high mobility, and, in some cases, radiological constraints. While some existing analytical techniques can detect and measure H and/or Li, they typically require extensive sample preparation, laboratory-based instrumentation, or extended analysis times that preclude real-time monitoring capabilities. To address these limitations, optical spectroscopy of laser-produced plasmas offers a promising solution, enabling field-deployable, standoff sensing of isotopic signatures with real-time analysis capability and no sample preparation requirement. This Topical Review summarizes fundamentals and recent advances in H and Li isotopic analysis using optical spectroscopy, with emphasis on laser-produced plasmas generated by laser ablation of solid targets. Key concepts relevant to isotope-resolved optical measurements are introduced, including isotope shifts, fine and hyperfine structure, and line-broadening mechanisms. Progress in optical emission, laser induced fluorescence, and absorption spectroscopy is reviewed for H and Li isotopes (\ce{^1H}, \ce{^2H}, \ce{^3H}, and \ce{^6Li}, \ce{^7Li}, respectively). The review concludes by identifying remaining challenges and opportunities for minimally destructive and field-deployable optical spectroscopy of light-element isotopes in complex environments.

\end{abstract}

\newpage
\tableofcontents

\section{Introduction}
\label{intro}

Light-element isotopes, particularly hydrogen (H) and lithium (Li), play a central role in modern science and technology as tracers of biological \cite{MishraIntJHealthSci2022}, geochemical \cite{PennistonRevMin2017}, and nuclear \cite{ChenePhilTrans2017, ShimadaBookChapter2020} processes. In planetary and hydrological systems, H isotopes trace the formation, transport, and phase evolution of water \cite{LecuyerChemGeo1998}, while Li isotopes record fractionation during fluid--rock interactions and weathering processes \cite{PistinerEPSL2003, TangIntGeoReview2007, TomascakBook2016}. In chemical and biological systems, isotope-dependent reaction rates provide insight into reaction pathways, transport phenomena, and kinetic isotope effects \cite{BasovMolecules2019, KlinmanAcctChemResearch2018, DiabateHealthPhysics1993, MishraIntJHealthSci2022}. In nuclear systems, H and Li isotopes are directly involved in fuel-cycle chemistry, tritium production, isotope inventory evolution, and reactor-material interactions, making their measurement important for safety, performance, and accountancy \cite{ShimadaBookChapter2020, ANLTechReport2020, deOliviera2017IAEA}.

Hydrogen exists as protium (\ce{^{1}H}, H), deuterium (\ce{^{2}H}, D), and radioactive tritium (\ce{^3H}, T). Protium accounts for 99.9885\% of natural H, while deuterium comprises about 0.0115\%. \ce{^3H}, with a half-life of 12.3~years, is produced naturally through cosmic-ray interactions in the upper atmosphere and is therefore present at trace levels in precipitation, surface water, and groundwater. Anthropogenic \ce{^3H} is generated through neutron-induced reactions in fission and fusion environments \cite{HernandezFusEngDesign2018, burns2012description}. Because \ce{^3H} has a low $\beta$-decay energy of 18.6~keV, it poses limited external radiation hazard but is a concern when incorporated into biological systems. If \ce{^3H} exists in the form of tritiated water, \ce{^{1}H^{3}HO}, it can readily exchange with ordinary water and enter organisms through ingestion, inhalation, or skin contact. \ce{^3H} may also become organically bound through metabolic reactions such as photosynthesis, leading to longer retention times and potentially elevated doses \cite{DiabateHealthPhysics1993}. 

Lithium has two stable isotopes, \ce{^6Li} and \ce{^7Li}, with natural abundances of approximately 7.5\% and 92.5\%, respectively. In addition to its two stable isotopes, Li has several short-lived radioactive isotopes from \ce{^4Li} to \ce{^13Li}, however, their rapid decay limits their relevance in most applied systems. Therefore, most Li isotopic analysis focuses on resolving and quantifying \ce{^{6}Li} and \ce{^{7}Li}. Lithium isotopes exhibit measurable fractionation during physical, chemical, and biological processes, making them powerful tracers in geochemistry and hydrology for studying weathering and fluid--rock interactions from the Earth's surface to the mantle \cite{ZhangNatureComm2022, PistinerEPSL2003, TangIntGeoReview2007}. Lithium isotopic measurements in mined and processed materials have also been demonstrated as 'fingerprints' for identifying the geographical location of Li, important in supply chain management of Li ion battery materials \cite{DesaultyNatureComm2022}. In engineered systems, Li isotope ratios are relevant to nuclear energy, defense technologies, and battery materials research. In nuclear systems, \ce{^6Li} and \ce{^7Li} are directly related to \ce{^3H} production \cite{HernandezFusEngDesign2018, burns2012description, ANLTechReport2020}. In Li ion batteries, studies of Li isotope redistribution before and after cycling can improve understanding of electrode degradation and changes in electrode-electrolyte phases impacting material performance \cite{ChangJPhysChemC2015, BattistellaACSEnergy2026}. 

\subsection{Challenges in measuring hydrogen and lithium isotopes}

The measurement of H and Li isotopes presents several analytical challenges. Hydrogen isotopes readily exchange with background species \cite{ParkNuclEngrTech2021} and exhibit high diffusivity and permeability in solids \cite{FukaiAdvPhys1985, UrrestizalaFusionEngr2023}, making sample integrity difficult to maintain between collection and analysis. Lithium isotopes can fractionate during sample preparation and processing, and measured ratios may be influenced by the chemical environment and matrix composition. For both elements, spatial heterogeneity in solid samples further complicates quantitative isotope analysis. These challenges are compounded by the limitations of conventional analytical techniques. 

Mass spectrometry is a highly matured technology, supported by the wide availability of commercial off‑the‑shelf instrumentation and it remains the benchmark for high-precision isotopic measurements, offering excellent detection limits and isotope ratio accuracy. However, mass spectrometry tools typically require destructive sampling, vacuum environments, calibration using standard reference materials, and extensive sample preparation. These constraints limit applicability for rapid, online measurements, particularly in high temperature, radioactive, or otherwise inaccessible environments where direct sample access can be difficult. 

Optical spectroscopy offers a complementary approach for non-contact, minimally destructive, and potentially standoff or field-deployable isotope measurements. In particular, optical spectroscopy of laser-produced plasmas (LPPs) has emerged as a promising route for analyzing H and Li isotopes directly from solids \cite{HariAPR2018, HarilalRMP2022, KurniawanAS2014}. LPPs generated by pulsed laser ablation (LA) produce transient plumes containing neutral atoms, ions, and molecules that can be interrogated spectroscopically to extract isotopic information \cite{HariAPR2018}. The use of LA plumes as an atomic reservoir for optical isotopic analysis of H and Li has been investigated for several decades, with early studies employing emission spectroscopy for H isotope detection \cite{KurniawanJAP2005} and absorption-based approaches for studying Li isotopes in LPPs \cite{SmithSCAB1998}. These studies demonstrated the ability to distinguish isotope shifts in spectral data. Advances in laser sources, high-resolution spectrometers, time-gated detection, and spectral modeling have expanded the ability of optical diagnostics to resolve H and Li isotopic shifts, enabling applications where conventional mass spectrometric techniques are impractical. Compared to mass spectrometry, optical spectroscopic methods for isotopic analysis remain an emerging field, and continued research efforts are needed to fully realize their potential across diverse applications. 

\begin{figure}[h]
    \centering
    \includegraphics[width=1\linewidth]{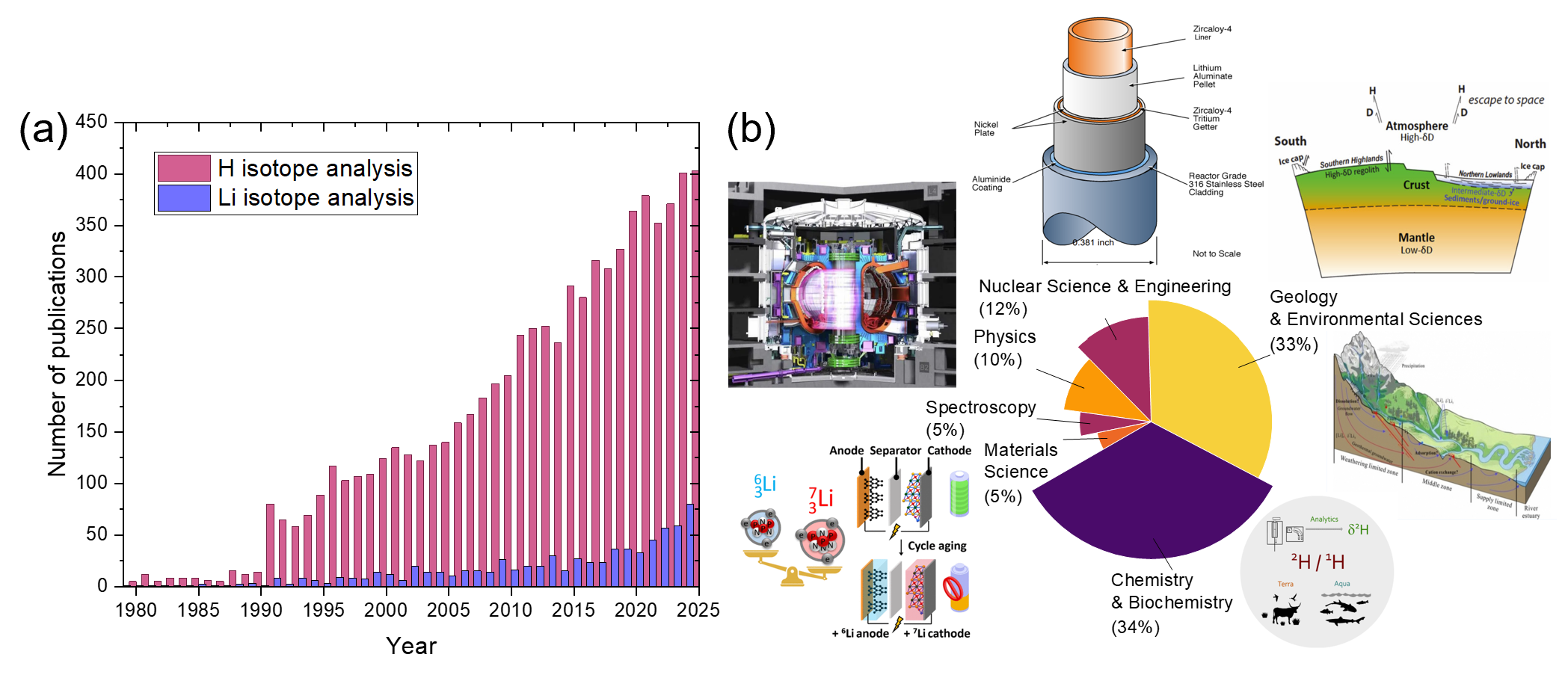}
    \caption{(a) Annual number of publications related to H and Li isotope analysis from 1980 to 2026, categorized by H-isotope studies (pink) and Li-isotope studies (purple). (b) Pie chart summarizing the disciplinary distribution of isotope-analysis publications, with corresponding figures illustrating application areas, with major contributions from Chemistry $\&$ Biochemistry, Geology $\&$ Environmental Sciences \cite{mars_water_reservoir_2024, jaimes2026COP}, Nuclear Science $\&$ Engineering \cite{aymar2002PPCF,  roy2025npj}, Physics, Materials Science \cite{BattistellaACSEnergy2026}, and Spectroscopy. Data used to generate this figure were collected from Web of Science (Clarivate) using the terms `hydrogen isotope analysis' and `lithium isotope analysis' as topic areas.}
    \label{fig:web_of_science}
\end{figure}

Accurate detection of Li and H isotopes is becoming increasingly important due to rapid advancements in advanced reactor, fusion and battery technologies. As an example, publication trends from 1980 to 2026 are shown in Figure~\ref{fig:web_of_science}(a), illustrating sustained and growing interest in H and Li isotope studies employing various measurement modalities. The disciplinary distribution shown in Figure~\ref{fig:web_of_science}(b) highlights the breadth of this work across chemistry and biochemistry, geology and environmental sciences, nuclear science and engineering, physics, materials science, and spectroscopy.

\subsection{Hydrogen and lithium isotopes in nuclear applications}

Hydrogen and Li isotopes are interconnected in nuclear energy systems, including both fission and fusion fuel cycles. Nuclear reactions in these environments can modify the isotopic composition of both H and Li species. A key example of this interconnection is the neutron-induced reaction of Li isotopes that generates tritium, directly linking Li and H isotope inventories. Several nuclear reactions relevant to H and Li isotopes in fission and fusion applications are summarized in Table~\ref{tab:HLi_reactions}. These reactions highlight a key concept: H and Li isotope inventories evolve dynamically in nuclear energy systems. This evolution necessitates continuous measurement and monitoring for various purposes (depending on the application), including material performance assessment and nuclear material accountancy.

\begin{table*}
\centering
\caption{Representative nuclear reactions relevant to H and Li isotope production, consumption, and monitoring in fusion and fission energy systems. Reactions are written with nuclide isotope notation with particle symbols defined as follows: $\beta^{-}$ is an emitted electron, and $\bar{\nu}_{e}$ is an electron antineutrino.}
\label{tab:HLi_reactions}
\renewcommand{\arraystretch}{1.3}
\footnotesize
\begin{tabularx}{\textwidth}{>{\raggedright\arraybackslash}p{4.8cm}
                                >{\raggedright\arraybackslash}p{3.6cm}
                                >{\raggedright\arraybackslash}X}
\hline
\textbf{Reaction} &
\textbf{System} &
\textbf{Comments} \\
\hline

\multicolumn{3}{l}{\textbf{Fusion energy applications}} \\
\hline

$\prescript{2}{1}{\mathrm{H}}
+\prescript{3}{1}{\mathrm{H}}
\rightarrow
\prescript{4}{2}{\mathrm{He}} + \prescript{1}{0}{\mathrm{n}}$
&
D--T fusion plasma
&
Primary energy-producing fusion reaction; requires tritium breeding.
\\

$\prescript{6}{3}{\mathrm{Li}}
+\prescript{1}{0}{\mathrm{n}}
\rightarrow
\prescript{3}{1}{\mathrm{H}}
+\prescript{4}{2}{\mathrm{He}}$
&
Tritium breeder blankets
&
Primary tritium breeding pathway; directly links $^{6}$Li enrichment to tritium production.
\\

$\prescript{7}{3}{\mathrm{Li}}
+\prescript{1}{0}{\mathrm{n}}
\rightarrow
\prescript{3}{1}{\mathrm{H}}
+ \prescript{4}{2}{\mathrm{He}} + \prescript{1}{0}{\mathrm{n}}$
&
Tritium breeder blankets
&
Secondary tritium production pathway with a neutron energy threshold.
\\

$\prescript{3}{1}{\mathrm{H}}
\rightarrow
\prescript{3}{2}{\mathrm{He}}
+\beta^{-}
+\bar{\nu}_{e}$
&
Fusion fuel cycle; tritium storage and handling systems
&
Tritium decay, inventory management, and fuel-cycle accountancy.\\

\hline
\multicolumn{3}{l}{\textbf{Fission energy applications}} \\
\hline

$\prescript{6}{3}{\mathrm{Li}}
+\prescript{1}{0}{\mathrm{n}}
\rightarrow
\prescript{3}{1}{\mathrm{H}}
+\prescript{4}{2}{\mathrm{He}}$
&
Li-containing molten salts; advanced fission systems
&
Major tritium production pathway in Li-containing salts when $^{6}$Li is present.
\\

$\prescript{7}{3}{\mathrm{Li}}
+\prescript{1}{0}{\mathrm{n}}
\rightarrow
\prescript{3}{1}{\mathrm{H}}
+\prescript{4}{2}{\mathrm{He}}+\prescript{1}{0}{\mathrm{n}}$
&
Li-containing molten salts; advanced fission systems
&
Fast-neutron tritium production pathway in lithium-containing reactor systems.\\

$\prescript{235}{92}{\mathrm{U}}
+\prescript{1}{0}{\mathrm{n}}
\rightarrow
\mathrm{fission\ fragments}
+\prescript{3}{1}{\mathrm{H}}$
&
Thermal and fast fission reactors
&
Tritium is produced at low yield through ternary fission; reaction shown schematically.
\\

$\prescript{239}{94}{\mathrm{Pu}}
+\prescript{1}{0}{\mathrm{n}}
\rightarrow
\mathrm{fission\ fragments}
+\prescript{3}{1}{\mathrm{H}}$
&
Fast reactors; Pu-bearing fuels
&
Tritium is produced at low yield through ternary fission; reaction shown schematically.
\\

$\prescript{3}{1}{\mathrm{H}}
\rightarrow
\prescript{3}{2}{\mathrm{He}}
+\beta^{-}
+\bar{\nu}_{e}$
&
Coolant, cover gas, off-gas, and waste systems
&
Relevant to tritium inventory, radiological control, and environmental monitoring.\\
\hline
\end{tabularx}
\end{table*}

In fusion systems, the deuterium--tritium reaction,
\ce{^{2}H + ^{3}H -> ^{4}He + ^{1}n}, is the primary fuel cycle of interest because it provides a high fusion cross section at accessible plasma temperatures \cite{NevinsJFusEnergy1998}. While deuterium can be extracted from water, tritium is not found in usable quantities in nature; therefore, sustained tritium breeding is required for continuous reactor operation. This is accomplished by using the neutron produced in the deuterium--tritium fusion reaction to drive neutron-induced reactions in Li-containing breeder blanket materials. The dominant tritium-production pathway (at thermal neutron energies) is the neutron absorption reaction, \ce{^{6}Li + ^{1}n -> ^{3}H + ^{4}He}.

A secondary pathway is the reaction of \ce{^7Li} with a neutron to produce \ce{^3H}, \ce{^4He}, and another neutron, \ce{^{7}Li + ^{1}n -> ^{3}H + ^{4}He + ^{1}n}, which becomes more relevant at higher neutron energies. Because \ce{^6Li} has a much higher thermal neutron capture cross section than \ce{^7Li}, \ce{^6Li}-enriched breeder materials are commonly considered for improving tritium breeding efficiency \cite{HernandezFusEngDesign2018, burns2012description}. Similar \ce{^{6}Li}-based \ce{^3H}-production pathways are also used in defense applications, including tritium-producing burnable absorber rods, where \ce{^{6}Li}-enriched \ce{LiAlO_2} is irradiated in a light-water reactor and the generated tritium is retained in surrounding components \cite{burns2012description}.

In fission systems, \ce{^3H} can be produced through neutron interactions with fuels, moderators, and coolants. \ce{^3H} production is expected to be greater in some advanced reactor concepts than in currently operating pressurized water, boiling water, and CANDU reactors, particularly in molten salt reactors (MSRs) that use Li-, Be-, or F-containing salts \cite{ANLTechReport2020}. Important \ce{^3H} production pathways include neutron-induced reactions involving \ce{^6Li}, \ce{^7Li}, as well as low-yield ternary fission during neutron-induced fission of \ce{^{235}U}, \ce{^{239}Pu}, and \ce{^{233}U}, as summarized in Table~\ref{tab:HLi_reactions}. Together, these pathways contribute to reactor \ce{^3H} inventories that require monitoring for operational safety, fuel-cycle management, and environmental control. For MSR concepts, \ce{^3H} generation rates can exceed those in commercial light-water reactors by orders of magnitude, depending on salt chemistry, neutron spectrum, and reactor design \cite{ANLTechReport2020}. Once produced, \ce{^3H} may be transported as tritiated species, aerosols, gases, or volatile compounds into cover-gas and off-gas systems, where it must be managed prior to recirculation or release control \cite{AndrewsNuclEngrDesign2021}.

Across both fission and fusion systems, isotope monitoring is complicated by the high mobility, permeability, and chemical exchange behavior of H isotopes. Tritium can permeate structural materials, become trapped at defects and interfaces, or redistribute among coolant, blanket, off-gas, and waste streams. Lithium isotope ratios may also evolve during operation as \ce{^6Li} is consumed or as Li-bearing materials interact with neutron fields and reactor chemistry. Therefore, measuring H and Li isotope ratios is important for validating transport models, assessing breeder-material performance, supporting \ce{^3H} accountancy, and reducing environmental release risks. These needs motivate deployable, isotope-sensitive diagnostics that can operate under conditions where conventional laboratory-based mass spectrometry is difficult to apply.

\subsection{Scope}

This review surveys recent developments in optical spectroscopic techniques for H and Li isotope analysis, with emphasis on LPPs generated from solid targets and their relevance to nuclear energy and defense applications. Literature from geochemistry, hydrology, battery materials, and analytical spectroscopy is included where it provides transferable insight into isotope-shift resolution, line broadening, matrix effects, optical thickness, or quantitative isotope ratio analysis. Information is also provided for detection of hydrogen isotopes in gas-phase molecular species, since the detection methods are similar whether or not a LPP is used to generate the molecular species. Section~\ref{experimental} surveys established analytical methods for H and Li isotope detection, placing optical spectroscopy in the broader analytical context. Section~\ref{fundamentals} introduces the fundamental principles of optical spectroscopy relevant to isotopic analysis, and Section~\ref{isotope_shifts} describes the isotopic shifts and fine and hyperfine structure of key H and Li atomic transitions. Section~\ref{LPPs} presents LPPs as versatile atomic reservoirs for solid sampling, and Section~\ref{line_broadening} discusses the line-broadening mechanisms that govern isotopic resolution in LPPs. Sections~\ref{H_isotopic_analysis} and~\ref{Li_isotopic_analysis} survey optical emission, absorption, and fluorescence techniques for H and Li isotope measurements, respectively. The review concludes in Section~\ref{summary_outlook} with a synthesis of the current state of the art and an outlook on key challenges and opportunities for future research.

\section{Experimental approaches for hydrogen and lithium isotopic detection and analysis}
\label{experimental}

\subsection{Comparative overview of analytical methods}

A wide range of techniques are available for detecting and quantifying H and Li isotopes, including mass spectrometry, radiation detection, magnetic resonance, nuclear and ion-beam probes, and optical spectroscopy. Table~\ref{tab:light_isotope_methods} provides a comparative overview of established methods, highlighting differences in measurement capability, analysis time, isotopic applicability, and key strengths and limitations. Optical spectroscopy techniques are excluded here and discussed separately. The reported measurement capabilities represent approximate detectable or quantifiable analyte levels, and the selected references emphasize applications to H and Li isotope analysis.

\begin{table}[htbp]
\caption{Summary of established methods used for the detection and analysis of H and Li isotopes. Techniques are categorized by mass spectrometry, radiation detection, magnetic resonance, nuclear and particle probes. Measurement capability is provided for each technique, in addition to relevant references. Acronyms and abbreviations used in the table are defined as follows: 
TD--MS (Thermal Desorption--Mass Spectrometry), 
ICP-MS (Inductively Coupled Plasma--Mass Spectrometry), TIMS (Thermal Ionization Mass Spectrometry), GD--MS (Glow Discharge--Mass Spectrometry), RIMS (Resonance Ionization Mass Spectrometry), SIMS (Secondary Ion Mass Spectrometry), LG--SIMS (Large Geometry--SIMS), ToF--SIMS (Time-of-Flight--SIMS), NanoSIMS (Nanoscale SIMS), APT (Atom Probe Tomography), LSC (Liquid Scintillation Counting), NMR (Nuclear Magnetic Resonance), NRA (Nuclear Reaction Analysis), NR (Neutron Reflectometry), ERDA (Elastic Recoil Detection Analysis), ppb (parts per billion), ppt (parts per trillion), ppq (parts per quadrillion), Bq (Becquerel), $\mu$M (micromolar).}
\label{tab:light_isotope_methods}
\scriptsize
\renewcommand{\arraystretch}{1.2}

\begin{tabularx}{\textwidth}{
p{2.5cm}
c
c
c
X
p{1.3cm}
}
\hline
\textbf{Method} &
\textbf{Capability} & 
\textbf{H} &
\textbf{Li} &
\textbf{Comments} &
\textbf{Ref.} \\
\hline

\multicolumn{6}{l}{\textbf{\textit{Mass Spectrometry}}} \\

TD--MS &
ppb--ppt &
$\checkmark$ &
&
Sensitive for mobile H isotopes, can study trapping behavior; destructive &
\cite{VonThermochimicaActa2003,PoonJNM2008,TapiaHEnergy2018,AshidaJNM1984} \\

ICP-MS &
ppt--ppq (Li) &
&
$\checkmark$ &
Trace Li isotope measurements possible; requires dissolution or laser ablation &
\cite{TomascakChemGeo1999,MillotGeo2004} \\

TIMS &
fg &
&
$\checkmark$ &
High precision Li isotope measurements; low throughput, extensive sample preparation required &
\cite{ChanAnChem1987,AggarwalAnMeth2016,ArienzoWater2020} \\

GD--MS &
ppb &
$\checkmark$ &
$\checkmark$ &
Well suited for bulk metallic samples; requires conductive specimens &
\cite{DonohueAnalyticalChem1991,BattistellaACSEnergy2026} \\

RIMS &
ppt &
$\checkmark$ &
$\checkmark$ &
High isotopic selectivity; complex instrumentation, limited availability. &
\cite{LevineMassSpec2009,SuryanarayanaMassSpec1998,YorozuAPB1999} \\

SIMS \textit{(LG, ToF, Nano)} &
ppm--ppb &
$\checkmark$ &
$\checkmark$ &
Spatially resolved isotope mapping with high sensitivity; nanoscale resolution possible with Nano-SIMS; destructive, reduced throughput for high-precision measurements. &
\cite{CarlsonAPL1978,StevieJVacSci2016,PennistonRevMin2017,DennyChemGeo2024,ZhuSIA2012,ZhaoACSAMI2023,BerthaultJPhysChemC2021,AbouraAppSS2021,LiJAAS2023} \\

APT &
ppm &
$\checkmark$ &
$\checkmark$ &
3D near atomic scale mapping; limited analysis volume, complex specimen preparation, H isotope detection difficult &
\cite{ChenMM2023,GaultMM2024,DevarajMatChar2021} \\

\hline
\multicolumn{6}{l}{\textbf{\textit{Radiation Detection (Tritium Only)}}} \\

LSC, Autoradiography, Proportional Counting &
Bq &
$\checkmark$ &
&
Sensitive detection of \ce{^3H}; activity-based measurement, not suitable for stable isotope ratio analysis &
\cite{BrayAnBiochem1960,VarlamAppRad2009,HuangJEnvRad2014,TuoJNucSci2008,ShmaydaFusionSciTech2002,SanadaJNuclSciTech2024,RanderathAnBiochem1970,LimaBiophysBiochem1962,ZouevFusionEngDes2000,OtsukaJACs2007} \\

\hline
\multicolumn{6}{l}{\textbf{\textit{Magnetic Resonance}}} \\

NMR &
$\mu$M &
$\checkmark$ &
$\checkmark$ &
Chemical-state specificity and non-destructive measurement; limited isotope ratio precision and sensitivity &
\cite{RaynesMolecPhys1971,BohmerProgNucMag2007} \\

\hline
\multicolumn{6}{l}{\textbf{\textit{Nuclear and Ion Beam Methods}}} \\

NRA / ERDA &
ppm &
$\checkmark$ &
$\checkmark$ &
Depth-resolved isotope profiling; requires ion-beam facilities and specialized infrastructure &
\cite{WangNuclInst2013,LaursenJNM1986,VykhodetsJETP2018,PretoriusNuclInst1988,SawickiNuclInst1986,TaylorAIPConf2019,TrocellierEAC2008} \\

NR &
at.\% &
$\checkmark$ &
$\checkmark$ &
Non-destructive interface sensitivity; requires access to neutron sources &
\cite{MunterPRB1997,GuascoNatureComm2022,HuegerJElectroChem2015,GaoNSR2023,HuegerJElectroChem2015} \\
\hline
\end{tabularx}
\end{table}

Mass spectrometry remains the benchmark for high-precision isotopic measurements, offering low detection limits and excellent isotope ratio accuracy. A wide range of mass spectrometric techniques are applicable to H and/or Li isotope analysis, although their suitability depends on the specific isotope and measurement approach. For example, thermal desorption mass spectrometry (TD–MS) is well established for H isotope analysis in materials where H is mobile, such as steels \cite{ChikadaFusEngr2009}, but is not applicable to Li isotopes. In contrast, inductively coupled plasma mass spectrometry (ICP–MS), particularly multi-collector (MC)-ICP-MS \cite{MillotGeo2004}, and thermal ionization mass spectrometry (TIMS) \cite{ArienzoWater2020} provide highly precise Li isotope measurements but are generally unsuitable for H analysis.

Several other techniques, including glow discharge mass spectrometry (GD–MS), resonance ionization mass spectrometry (RIMS), secondary ion mass spectrometry (SIMS), and atom probe tomography (APT), can resolve both H and Li isotopes. GD–MS enables bulk compositional and isotopic analysis through plasma-assisted sputtering and ionization, particularly for conductive materials \cite{DonohueAnalyticalChem1991}. RIMS combines element-selective laser excitation with mass spectrometric detection, providing high isotopic selectivity via resonant ionization \cite{LevineMassSpec2009}. SIMS has a range of configurations, including large geometry SIMS (LG-SIMS), time of flight SIMS (ToF-SIMS), and NanoSIMS, each offering high mass resolving power and spatially resolved isotopic analysis \cite{VanSIMSBook2014, SenonerJAAS2012, StevieJVacSci2016, DennyChemGeo2024, ZhuSIA2012, ZhaoACSAMI2023, BerthaultJPhysChemC2021, AbouraAppSS2021, LiJAAS2023}. APT, while often categorized as a microscopy tool, functions as a specialized time-of-flight mass spectrometer capable of three-dimensional, near-atomic-scale compositional and isotopic mapping \cite{GaultNature2021}. Although APT has been applied to both Li \cite{DevarajMatChar2021} and H isotopes \cite{GaultMM2024, ChenMM2023}, quantitative H isotope analysis remains challenging due to residual chamber H and limited mass resolving power relative to other techniques. Despite their analytical capabilities, most mass spectrometric techniques require high vacuum environments and extensive or destructive sample preparation, limiting throughput, portability, and real-time applicability in dynamic or harsh environments.

For \ce{^3H}, radiation-based techniques provide a complementary detection strategy by directly measuring low-energy $\beta$ particles emitted during \ce{^3H} decay. Methods such as liquid scintillation counting (LSC), gas proportional counting, surface scintillation, and autoradiography enable sensitive quantification through detection of ionization or scintillation events \cite{BrayAnBiochem1960, TuoJNucSci2008, ShmaydaFusionSciTech2002, LimaBiophysBiochem1962}. While these approaches are essential for environmental monitoring, regulatory compliance, and nuclear fuel-cycle safeguards \cite{HuangJEnvRad2014, SanadaJNuclSciTech2024}, they are inherently limited to radioactive isotopes and are not applicable to stable H or Li isotopes. In addition, the low energy of \ce{^3H} $\beta$ decay (18.6~keV) restricts standoff detection capabilities \cite{parkerelesevier2023}, confining their use to specific systems and geometries \cite{theodorsson1999review}.

Magnetic resonance techniques, including nuclear magnetic resonance (NMR), provide non-destructive insight into local bonding environments, diffusion behavior, and chemical states of H and Li isotopes in condensed matter \cite{HoreNMRBook2015}. NMR is sensitive to nuclei such as \ce{^1H}, \ce{^2H} \cite{RaynesMolecPhys1971}, \ce{^6Li}, and \ce{^7Li} \cite{BohmerProgNucMag2007, BerthaultPCCP2023}, enabling isotopically specific measurements and, in some cases, simultaneous multi-nuclear analysis. However, NMR suffers from inherently low sensitivity, long acquisition times, and reliance on high-field superconducting magnets, limiting portability and real-time applicability. Consequently, its role in quantitative isotope analysis is largely confined to controlled laboratory environments.

Nuclear and ion-beam techniques, including nuclear reaction analysis (NRA), neutron reflectometry (NR), and elastic recoil detection analysis (ERDA), enable quantitative, depth-resolved measurements of light elements through interactions of energetic ions or neutrons with target nuclei. These methods provide ppm-level sensitivity with nanometer-scale depth resolution and have been widely applied to H and, to a lesser extent, Li isotopes. However, they require accelerator or neutron facilities, vacuum conditions, and specialized sample preparation, restricting their use to dedicated laboratory environments and precluding field deployment.


In contrast to these predominantly laboratory-based approaches, spectroscopy methods offer opportunities for rapid, minimally destructive, and potentially fieldable and standoff-capable isotope analysis. Here, the term 'standoff' refers to measurement geometries in which signals from a target can be collected without direct contact, such that no physical access to, collection of, or preparation of the sample is required. This definition has been used previously \cite{HariAPR2018, MunsonBookChapter2011} in the context of laser- and spectroscopy-based diagnostics for standoff detection of hazardous and radiological materials \cite{GottfriedAnBioChem2009, GaonaSCAB2014}. Spectroscopy methods capable of standoff, online or in situ measurements include optical techniques in the UV–VIS–NIR range (e.g., LIBS, LAS, LIF, Raman) as well as infrared methods such as Fourier Transform Infrared Spectroscopy (FTIR) and LAS using IR laser sources. Such approaches complement conventional techniques by enabling rapid analysis with minimal sample preparation and, in some cases, remote or in situ measurements \cite{HariAPR2018, GaftOM2008, PachecoABC2009}.

\subsection{Optical spectroscopy of laser-produced plasmas}

LPPs generated through LA create transient, high temperature environments that facilitate the formation of excited atomic and molecular species. The optical emission and absorption from these plasmas provides access to multiple isotope-sensitive spectral features, discussed in Section \ref{fundamentals}. The temporal evolution of LPP emission and absorption properties, from initial continuum radiation, to ionized and neutral atomic species, to molecular formation during plasma cooling, enables selective observation of different isotopic signatures with time-gated detection \cite{HariAPR2018}. Furthermore, the localized nature of LA (typically tens to hundreds of $\mu$m spot size) allows for spatially resolved isotopic mapping, while the standoff capability of laser systems enables remote analysis.

Hence, trade-offs exist between approaches summarized in Table~\ref{tab:light_isotope_methods} and spectroscopy tools, with regard to analytical performance, measurement time, and deployability. Techniques offering the highest isotopic precision typically rely on destructive sampling, vacuum environments, or specialized infrastructure, whereas approaches capable of in situ or standoff measurements often face limitations related to measurement sensitivity. Within this context, optical spectroscopy of LPPs occupies a unique position by combining standoff capability, temporal and spatial resolution, and access to isotopic information, enabling real-time, minimally destructive isotopic analysis with the potential for both laboratory and field applications. 

Another advantage of optical spectroscopy is that, unlike radiation-detection methods, it can use \ce{^2H} as a nonradioactive spectroscopic and chemical surrogate for \ce{^3H}. Method development performed using \ce{^2H} can therefore be extended to more costly and hazardous \ce{^3H} measurements, provided that the relevant \ce{^3H} spectroscopic transitions are known or can be measured.

However, despite these advantages, practical implementation remains an important consideration. In contrast to many methods listed in Table~\ref{tab:light_isotope_methods}, high-resolution optical spectroscopy tools for isotope analysis in LPPs are not generally available as commercial off-the-shelf systems. Commercial LIBS instruments exist, but isotope-resolved measurements often require custom experimental set-ups including high-resolution spectrographs \cite{Smith2002, CremersAS2012, caleb2026}. No commercial system is available for LIF and LAS, which currently use a range of custom experimental designs and implementations. Several commercial laboratory and portable instruments are available for analysis of isotopes in gas-phase molecular species using tunable laser absorption spectroscopy, suggesting that LPP-based instrumentation may also be feasible in the future.

\section{Fundamentals of optical spectroscopy for isotopic analysis}
\label{fundamentals}

Optical spectroscopy provides a foundational framework for probing atoms and molecules in a system, either solid, liquid, or gas, depending on the technique and sampling approach. Atoms and molecules undergo transitions between discrete, quantized energy levels by absorbing or emitting light at characteristic wavelengths or frequencies. The resulting spectral signatures encode information about the species present, the thermodynamic state of the medium, and the underlying atomic or molecular structure. For isotope analysis, these same transitions also contain subtle signatures of nuclear mass and charge that manifest as measurable frequency (or wavelength) shifts between isotopes of the same element.

The frequency or wavelength of a spectral line is determined by the energy difference between the upper and lower electronic states:
\begin{equation}
E_2 - E_1 = \frac{hc}{\lambda} = h\nu,
\label{eq:energy_transition}
\end{equation}
where $h$ is Planck’s constant, $c$ is the speed of light, $\lambda$ is the photon wavelength, and $\nu$ is the corresponding frequency. Thus, the optical spectrum directly reflects the electronic structure of the species participating in the transition.

The emission intensity and shape of an observed spectral line depend on the population of excited states and the likelihood of radiative transitions. The spectral emission coefficient for spontaneous emission of a transition from level 2 to level 1 at frequency $\nu$ is:
\begin{equation}
\epsilon(\nu)_{2\rightarrow1} = \frac{h\nu}{4\pi} A_{21} N_2 \Phi_\nu,
\end{equation}
where $A_{21}$ is the Einstein coefficient for spontaneous emission, $N_2$ is the excited-state population, and $\Phi_\nu$ represents the normalized line-shape function. Other mechanisms contributing to lineshapes and linewidths are discussed in Section \ref{line_broadening}. The absorption coefficient for a transition from level 1 to level 2 is:
\begin{equation}
\alpha(\nu)_{1\rightarrow2} = 
\frac{h\nu}{4\pi} (N_1 B_{12} - N_2 B_{21}) \Phi_\nu,
\end{equation}
with $B_{12}$ and $B_{21}$ the Einstein coefficients for stimulated absorption and emission.

Under local thermodynamic equilibrium (LTE), the population of states follows a Boltzmann distribution:

\begin{equation}
n_i = \frac{g_i n_{\mathrm{tot}}}{U(T)}
\exp\left(-\frac{E_i}{k_B T_{\mathrm{ex}}}\right),
\end{equation}

\noindent where $n_i$ is the population density of the $i$-th energy level, $g_i$ is the statistical weight or degeneracy of that level, $n_{tot}$ is the total number density of the species, $U(T)$ is the temperature-dependent partition function, $E_i$ is the energy of the $i$-th level, $k_B$ is the Boltzmann constant, and $T_{ex}$ is the excitation temperature. This relationship links measured emission intensity or absorbance to the state population, enabling plasma temperature, species density, and composition to be inferred when LTE assumptions are valid. The existence of LTE in LPP requires that the electron-atom and electron-ion collisional processes occur sufficiently fast and dominate the radiative processes, as discussed elsewhere \cite{HarilalRMP2022, CristoforettiSCAB2010}. However, it is worth noting that energy level populations for particular species may follow Boltzmann distributions (although not necessarily at the same temperatures) even in the absence of this strict LTE definition, as observed for several absorption spectroscopy measurements \cite{2021-SCAB-Weeks, phillips2023comparison, wala2025characterization}. 

Molecular spectra involve added complexity due to vibrational and rotational degrees of freedom. The total energy of a rovibronic state can be written as:
\begin{equation}
E(e,v,J) = T_e + G_v + F_J,
\end{equation}
where $T_e$ is the electronic contribution, $G_v$ the vibrational energy, and $F_J$ the rotational term. 

Molecular isotope effects are primarily governed by changes in the reduced mass of the molecule ($\mu$ = $(m_1  m_2)/(m_1+m_2)$), which directly influence vibrational and rotational energy spacings. To first order, vibrational frequencies scale as $\nu_v \propto \sqrt{k/\mu}$ and rotational constants as $B \propto 1/\mu$, where $\mu$ is the reduced mass and $k$ is the effective bond stiffness. As a result, isotopic substitution produces relatively large shifts in vibrational and rotational structure compared to atomic electronic transitions. These shifts often manifest as shifts in band heads or changes in rotational spacing within molecular spectra, allowing molecular spectroscopy to be used for isotopic detection.

However, spectroscopy of molecules in LPPs is often limited by the transient and high temperature nature of the plasma, where molecular species may dissociate or exist only at later stages of plasma evolution. Additionally, molecular spectra consist of dense, overlapping rovibrational bands that can complicate spectral features, particularly when Doppler broadening can be significant. Consequently, while molecular transitions can offer isotope sensitivity, atomic transitions have been more commonly employed in LPP-based diagnostics.

In summary, optical spectroscopy provides direct access to atomic and molecular energy levels, and optical transitions can serve as indicators of plasma physical conditions, plasma species present, and information about elements and isotopes present in the target material of interest. While the fundamental structure of these transitions is governed by electronic, vibrational, and rotational energies, isotope-dependent changes arise from differences in nuclear mass and charge distribution. These nuclear effects, which shift the electronic energy levels and therefore the atomic spectral line or molecular band positions, are described in the following Section \ref{isotope_shifts}.

\section{Isotopic shifts in hydrogen and lithium spectral data}
\label{isotope_shifts}
Isotope shifts are small differences in the electronic transition energies of atoms that arise for isotopes of the same element. These shifts originate from nuclear-level properties, specifically the finite nuclear mass and the spatial distribution of nuclear charge \cite{king2013isotope}. The resulting energy differences manifest as subtle yet measurable wavelength or frequency shifts in optical spectra, forming the basis of isotope-selective optical diagnostics. Two primary mechanisms contribute to isotope shifts, namely: (1) mass shift (MS), and (2) field shift (FS). Together, these contributions modify the electronic potential and alter the observed transition wavelengths.

In the case of MS, the transition energy varies for atoms that have the same nuclear and electronic charges but differ in nuclear mass. Consequently, the energy of the transition varies among different isotopes. The MS can be further categorized into two components: normal mass shift (NMS) and specific mass shift (SMS). NMS is due to the introduction of the reduced mass ($\mu$) of the electron-nuclear system in kinetic energy. In this case, at the first order, the isotope-dependent correction to an energy level $(n, j)$ is expressed as:
\begin{equation}
\Delta E_{n j}^{A A'} =
\left( \frac{m_e}{M_A} - \frac{m_e}{M_{A'}} \right) 
E_{n j},
\label{eq:isotope_energy_shift}
\end{equation}
where $M_A$ and $M_{A'}$ are the nuclear masses of the isotopes.

SMS arises from correlations in the motion of electrons and the influence of those correlations on the nuclear recoil energy, and can be expressed as:
\begin{equation}
\Delta E_{\mathrm{SMS}}^{A A'} 
= K_{\mathrm{SMS}}
\left( \frac{1}{M_{A'}} - \frac{1}{M_A} \right),
\label{eq:specific_mass_shift}
\end{equation}
where $\Delta E_{\mathrm{SMS}}^{A A'}$ is the specific mass shift in transition energy between isotopes with mass numbers $A$ and $A'$, $M_A$ and $M_{A'}$ are the corresponding nuclear masses, and $K_{\mathrm{SMS}}$ is the specific mass shift constant. The constant $K_{\mathrm{SMS}}$ is given by:
\begin{equation}
K_{\mathrm{SMS}} = 
\left\langle
\sum_{n<k} \hat{\boldsymbol{p}}_n \cdot \hat{\boldsymbol{p}}_k
\right\rangle,
\end{equation}
where $\hat{\boldsymbol{p}}_n$ and $\hat{\boldsymbol{p}}_k$ are the momentum operators of the $n$th and $k$th electrons, respectively, and the summation is taken over all unique electron pairs. The angle brackets denote the expectation value over the electronic state of interest. The SMS is generally smaller than the NMS.

The \textit{field shift} originates from differences in the finite size and charge distribution of the nucleus between isotopes. The correction is the following:
\begin{equation}
\Delta E_{\mathrm{FS}}^{A A'} 
= 
F_i\,\Delta \langle r^2 \rangle^{A A'},
\end{equation}
where $F_i$ is the field-shift constant and $\Delta \langle r^2 \rangle^{A A'}$ is the difference in mean-square charge radii.  

The total isotope shift for level $i$ is:
\begin{equation}
\Delta E_i^{A A'} 
= 
\frac{M_{A'} - M_A}{M_A M_{A'}} K_i 
+
F_i \Delta \langle r^2 \rangle^{A A'},
\label{eq:total_isotope_shift}
\end{equation}
where $K_i$ includes both NMS and SMS contributions. 

In lighter atoms, such as H and Li, isotope shifts are predominantly governed by the mass shift because of the relatively large fractional difference in nuclear mass between isotopes \cite{nortershauser2011isotope}. Conversely, in heavier atoms, such as U and Pu, the field shift arising from differences in nuclear charge distribution becomes the dominant contribution \cite{HariAPR2018}. As a result, optical isotope analysis has been most successful for light elements and selected heavy elements, whereas the much smaller isotope shifts of most intermediate-mass elements remain difficult to resolve with optical spectroscopy of LPPs.

In molecular transitions, isotope shifts result from changes in rotational and vibrational constants which depend on the atomic masses. In a harmonic oscillator approximation, a diatomic molecule between atoms with masses $m_1$ and $m_2$ has quantized vibrational energy levels given by:

\begin{equation}
E_v = \hbar \sqrt{\frac{k}{\mu}} \cdot (v+1/2)
\end{equation}

\noindent where $k$ is the force constant, $v$ is the vibrational quantum number. Likewise, in a rigid rotor approximation of a diatomic molecule, the rotational energy levels are given by:

\begin{equation}
E_J = hcBJ(J+1) 
\end{equation}

\noindent where the rotational constant $B$ depends on the reduced mass $\mu$ and bond length $r$ via

\begin{equation}
B = \frac{h}{8\pi^2 c \mu r^2} 
\end{equation}

In real molecules, higher-order perturbations are needed to describe the rotational and vibrational energy levels; however, the harmonic oscillator and rigid rotor models illustrate how isotopic substitutions shift the associated energy levels through the reduced mass, leading to corresponding shifts in vibronic (electronic+vibrational+rotational), rovibrational (vibrational+rotational), and pure rotational transitions. While more complicated to model, transitions in polyatomic molecules also experience isotope shifts.

The atomic transitions of H and Li exhibit fine structure, which originates from spin–orbit coupling and split states with identical $n$ and $l$ but different total electronic angular momentum J. Each transition may also exhibit hyperfine structure (hfs), arising from the interaction between nuclear dipole moments and the electromagnetic fields generated by electrons near the nucleus. When the nucleus possesses spin angular momentum (I), it can couple with the total electronic angular momentum (J) to produce the total angular momentum of the atom (F). This coupling gives rise to hfs, which appear as components or transitions between different hyperfine levels \cite{van2013advances, DasPRA2007, HarilalRMP2022}. In H and Li, the hyperfine splittings are small relative to isotope shifts, but can be important for accurate spectral modeling.  
 
The reported IS of prominent transitions in H and Li within the ultraviolet and visible spectral range are summarized in Tables \ref{tab:H_atomic_isotope_shifts} and \ref{tab:Li_isotope_shifts}. In Table \ref{tab:H_atomic_isotope_shifts}, H isotope atomic transitions are listed, including \ce{^1H}/\ce{^2H} and select examples of \ce{^2H}/\ce{^3H}. The IS values for \ce{^1H}/\ce{^2H} are relatively high, ranging from $\approx$100 - 180 pm. Among these transitions, H$_\alpha$ at 656.28~nm exhibits the largest isotopic shift (180 pm) and is also the strongest transition in the Balmer series due to its relatively high transition probability \cite{NIST_database}. Spectral broadening also varies among the Balmer transitions. H$_\alpha$ exhibits less Stark broadening than higher-order Balmer transitions such as H$_\beta$, which involve higher principal quantum numbers and are more sensitive to electron density dependent Stark broadening. The combination of a large isotopic shift, strong emission intensity, and reduced spectral broadening makes H$_\alpha$ well suited for H isotopic analysis in the VIS spectral region \cite{KurniawanAS2014,KautzOE2021,BurgerPoP2018}. The \ce{^2H}/\ce{^3H} IS is smaller than that of \ce{^1H}/\ce{^2H}. Although not explicitly listed here, the IS for \ce{^1H}/\ce{^3H} has been experimentally measured, and is $\approx$240 pm for H$_\alpha$ \cite{HarilalJAAS2024}. 

\begin{figure}[h]
\centering
\includegraphics[width=0.55\linewidth]{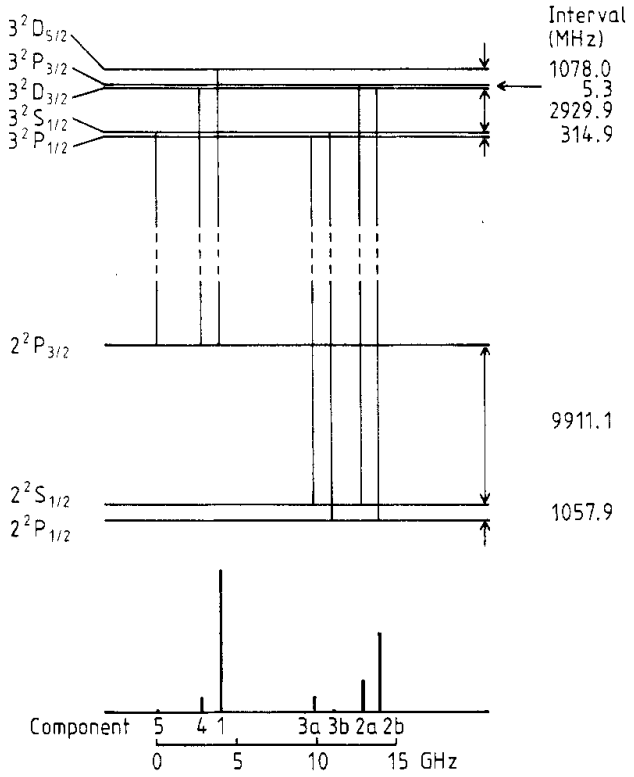}
\caption{\label{fig:H_energylevels} Energy level diagram and transition components for the Balmer $H_{\alpha}$ transition. The diagram shows fine-structure levels. The splittings (in MHz) denote hyperfine splitting for tritium, and the vertical lines indicate allowed optical transitions between ground and excited states. The lower panel presents the relative frequency positions of the resolved hyperfine components (labeled 1--5, from highest to lowest intensity) on a GHz scale, illustrating the spectral structure \cite{TateJPhysB1988}.} 
\end{figure}

\begin{figure}[h]
\centering
\includegraphics[width=0.75\linewidth]{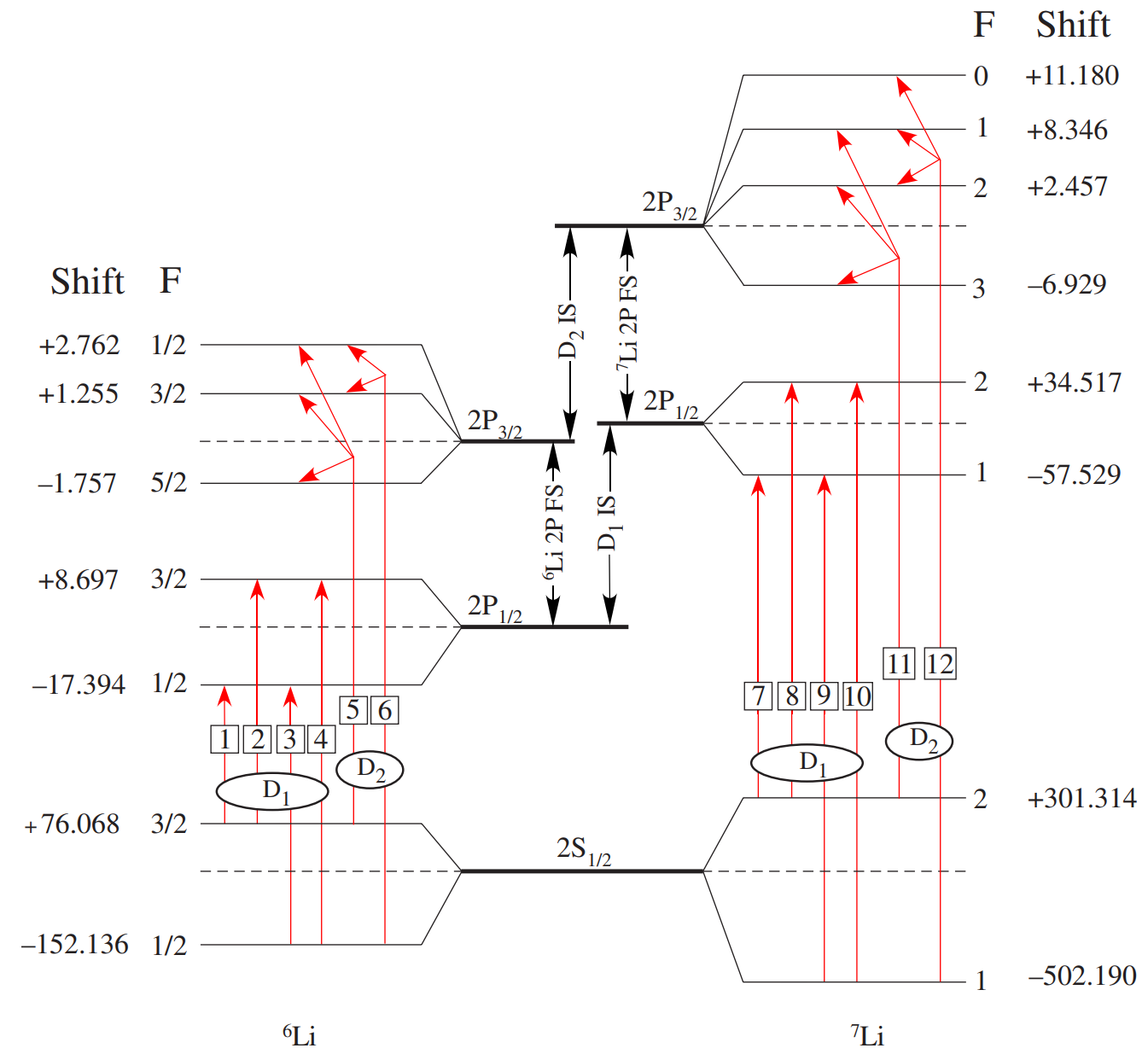}
\caption{\label{fig:Li_energylevels} Relevant fine structure (FS) and hyperfine energy levels in $^6$Li and $^7$Li. Each hyperfine level is labeled with the value of its quantum number $F$, and the shift of the level (in MHz) from the center of gravity of the state is given alongside. The fine structures are marked as D$_1$ and D$_2$. The various hyperfine transitions for $^6$Li are labeled from 1 to 6 and for  $^7$Li are labeled from 7-12. Transitions in the D$_2$ line are not fully resolved and are shown with the multiple hyperfine levels that couple to a given ground level. The tip of the vertical line indicates the approximate location of the peak center \cite{DasPRA2007}.} 
\end{figure}

For Li, Table \ref{tab:Li_isotope_shifts} summarizes several transitions and their corresponding isotopic shifts in the UV-VIS-NIR spectral region \cite{RadziemskiPRA1995}. An important trend observed in Table \ref{tab:Li_isotope_shifts} is that the majority of IS are from $\approx$ 2 - 5 pm, with only one IS with a greater value of $\approx$15.8 pm (Li I centered around 670.8 nm). Hence the H$_\alpha$ and the Li I 670.8 nm transitions are regularly utilized for isotope analysis through optical spectroscopy. Energy level diagrams are given in Figures \ref{fig:H_energylevels} and \ref{fig:Li_energylevels} for 
the Balmer $H_{\alpha}$ and Li I 670.8 nm transitions, respectively. 

\begin{table}[h!]
\centering
\caption{Summary of atomic transitions used in H isotopic analysis. Wavelength and isotopic shift data for \ce{^{1}H}/\ce{^{2}H} are summarized from Reference \cite{NIST_database}. Isotopic shifts refer to the difference in wavelength between H isotopes. Wavelengths below are air wavelengths. The \ce{^{2}H}/\ce{^{3}H} shift for \ce{H_{\alpha}} is reported from simulated data given in Reference \cite{VujadinovicEPhys2026}.}
\label{tab:H_atomic_isotope_shifts}

\renewcommand{\arraystretch}{1.25}
\setlength{\tabcolsep}{3pt}
\footnotesize

\begin{tabularx}{\linewidth}{
    >{\centering\arraybackslash}p{1.4cm}
    >{\centering\arraybackslash}X
    >{\centering\arraybackslash}X
    >{\centering\arraybackslash}X
    >{\centering\arraybackslash}X
    >{\centering\arraybackslash}X}
\hline
\textbf{Transition} &
\textbf{\ce{^{1}H} wavelength (nm)} &
\textbf{\ce{^{2}H} wavelength (nm)} &
\textbf{\ce{^{3}H} wavelength (nm)} &
\textbf{\ce{^{1}H}/\ce{^{2}H} IS (pm)} &
\textbf{\ce{^{2}H}/\ce{^{3}H} IS (pm)} \\
\hline

H\textsubscript{$\alpha$}   & 656.279 & 656.099 & 656.039 & $-180$ & $-60$ \\
H\textsubscript{$\beta$}    & 486.135 & 486.000 & 485.955 & $-135$ & $-45$ \\
H\textsubscript{$\gamma$}   & 434.047 & 433.928 &         & $-119$ &       \\
H\textsubscript{$\delta$}   & 410.171 & 410.062 &         & $-109$ &       \\
H\textsubscript{$\epsilon$} & 397.008 & 396.899 &         & $-108$ &       \\
H\textsubscript{$\zeta$}    & 388.906 & 388.799 &         & $-107$ &       \\
H\textsubscript{$\eta$}     & 383.540 & 383.434 &         & $-106$ &       \\

\hline
\end{tabularx}
\end{table}

\begin{table*}[htbp]
\centering
\caption{Summary of \ce{Li I} transitions, including wavelengths and isotopic shifts for the UV-VIS-NIR spectral region. Wavenumber values and IS in \ce{cm^{-1}} are reproduced from Reference \cite{RadziemskiPRA1995}. Vacuum wavelength was converted from wavenumber values, and air wavelength was converted from vacuum wavelength from Reference \cite{NistCiddor2025}. Isotopic shift (IS) in pm is the difference in wavelength between \ce{^6Li} and \ce{^7Li} wavelengths, and converted to pm.}
\label{tab:Li_isotope_shifts}
\small
\setlength{\tabcolsep}{4pt}
\begin{tabular}{cccccccc}
\toprule
\multicolumn{2}{c}{Wavenumber} &
 &
\multicolumn{2}{c}{Wavelength (vacuum)} &
\multicolumn{2}{c}{Wavelength (air)} &
 \\
\cmidrule(lr){1-2} \cmidrule(lr){4-5} \cmidrule(lr){6-7}
\ce{^{6}Li} & \ce{^{7}Li} &
IS &
\ce{^{6}Li} & \ce{^{7}Li} &
\ce{^{6}Li} & \ce{^{7}Li} &
IS \\
(\si{cm^{-1}}) & (\si{cm^{-1}}) & (\si{cm^{-1}}) &
(\si{nm}) & (\si{nm}) &
(\si{nm}) & (\si{nm}) & (\si{pm}) \\
\midrule
30925.08 & 30925.55 & \multirow{2}{*}{0.48}
& 323.362 & 323.357 & 323.271 & 323.266 & 4.9 \\
30925.17 & 30925.65 &
& 323.361 & 323.356 & 323.270 & 323.265 & 5.0 \\
\midrule
25533.03 & 25533.25 & \multirow{3}{*}{0.22}
& 391.650 & 391.646 & 391.541 & 391.537 & 3.4 \\
25533.03 & 25533.26 &
& 391.650 & 391.646 & 391.541 & 391.537 & 3.5 \\
25533.36 & 25533.59 &
& 391.644 & 391.641 & 391.536 & 391.532 & 3.5 \\
\midrule
25083.408 & 25083.610 & \multirow{2}{*}{0.201}
& 398.670 & 398.667 & 398.559 & 398.556 & 3.2 \\
25083.744 & 25083.945 &
& 398.665 & 398.661 & 398.554 & 398.551 & 3.2 \\
\midrule
24190.6880 & 24190.886 & \multirow{3}{*}{0.198}
& 413.382 & 413.379 & 413.268 & 413.264 & 3.4 \\
24190.6950 & 24190.894 &
& 413.382 & 413.379 & 413.268 & 413.264 & 3.4 \\
24191.0230 & 24191.221 &
& 413.376 & 413.373 & 413.262 & 413.259 & 3.4 \\
\midrule
23395.3000 & 23395.4805 & \multirow{2}{*}{0.1831}
& 427.436 & 427.433 & 427.318 & 427.315 & 3.3 \\
23395.6300 & 23395.8158 &
& 427.430 & 427.427 & 427.312 & 427.309 & 3.4 \\
\midrule
21719.1900 & 21719.3539 & \multirow{3}{*}{0.1624}
& 460.422 & 460.419 & 460.296 & 460.292 & 3.5 \\
21719.2100 & 21719.3691 &
& 460.422 & 460.419 & 460.295 & 460.292 & 3.4 \\
21719.5300 & 21719.6893 &
& 460.415 & 460.412 & 460.289 & 460.285 & 3.4 \\
\midrule
20107.9107 & 20108.0490 & \multirow{2}{*}{0.1383}
& 497.317 & 497.313 & 497.181 & 497.177 & 3.4 \\
20108.2460 & 20108.3843 &
& 497.308 & 497.305 & 497.172 & 497.169 & 3.4 \\
\midrule
16378.9728 & 16379.0661 & \multirow{3}{*}{0.0932}
& 610.539 & 610.535 & 610.373 & 610.370 & 3.5 \\
16379.0089 & 16379.1021 &
& 610.538 & 610.534 & 610.372 & 610.368 & 3.5 \\
16379.3081 & 16379.4014 &
& 610.526 & 610.523 & 610.361 & 610.357 & 3.5 \\
\midrule
14903.2973 & 14903.6483 & \multirow{2}{*}{0.3511}
& 670.992 & 670.977 & 670.811 & 670.795 & 15.8 \\
14903.6327 & 14903.9838 &
& 670.977 & 670.962 & 670.796 & 670.780 & 15.8 \\
\midrule
12302.0799 & 12302.1110 & \multirow{2}{*}{0.0311}
& 812.871 & 812.869 & 812.651 & 812.649 & 2.1 \\
12302.4152 & 12302.4463 &
& 812.849 & 812.846 & 812.629 & 812.627 & 2.1 \\
\bottomrule
\end{tabular}
\end{table*}


The energy level diagram presented in Figure ~\ref{fig:H_energylevels} illustrates the fine- and hyperfine-structure components of the Balmer H$_\alpha$ transition. These components were resolved using Doppler-free saturated absorption spectroscopy of atomic tritium (\ce{^3H}) in a discharge containing H isotopes, allowing MHz-level resolution of individual spectral components \cite{TateJPhysB1988}. The transition arises from multiple fine-structure levels that produce several allowed electric dipole transitions (labeled 1, 2a, 2b, 3a, 3b, 4, and 5, in order of decreasing intensity), each exhibiting isotope-dependent shifts relative to \ce{^1H} and \ce{^2H}. The numerical values (in MHz) shown in the figure represent the relative frequency offsets of these transitions with respect to a reference line center, corresponding to separations between fine- and hyperfine-resolved components. These spectral features span several orders of magnitude, including isotope shifts on the order of $\approx$160~GHz, fine-structure separations of $\approx$5~GHz, and hyperfine splittings ranging from tens to hundreds of MHz \cite{TateJPhysB1988}. The Doppler-free linewidths of $\approx$100~MHz enable resolution of larger hyperfine components, such as the $2S_{1/2}$ level splitting ($\approx$177~MHz in \ce{^1H} and $\approx$189~MHz in \ce{^3H}), while smaller splittings, such as those in \ce{^2H} ($\approx$27~MHz), remain unresolved \cite{TateJPhysB1988}. Therefore, Figure ~\ref{fig:H_energylevels} illustrates how multiple closely spaced transitions contribute to the observed spectral structure of the H$_\alpha$ line. More recently, Kramida \cite{KramidaAtomicData2010} compiled and evaluated high-precision spectroscopic measurements of \ce{^1H}, \ce{^2H}, and \ce{^3H} to construct updated energy-level datasets, fine- and hyperfine-structure intervals, and Ritz wavelengths.

The energy level diagram presented in Figure \ref{fig:Li_energylevels} illustrates the fine-structure and hyperfine energy levels of the Li I 670.8~nm resonance transition for both \ce{^6Li} and \ce{^7Li} \cite{DasPRA2007}. Each hyperfine level is labeled by the total angular momentum quantum number $F$, with the corresponding frequency shifts (in MHz) given relative to the center of gravity of each electronic state. The diagram shows the splitting of the $2S_{1/2}$ ground state and the $2P_{1/2}$ and $2P_{3/2}$ excited states into multiple hyperfine levels, resulting from nuclear spin interactions that differ between \ce{^6Li} and \ce{^7Li}. Allowed electric dipole transitions between these levels are indicated by vertical arrows, giving rise to multiple closely spaced spectral components associated with the D$1$ ($2S_{1/2} \rightarrow 2P_{1/2}$) and D$2$ ($2S_{1/2} \rightarrow 2P_{3/2}$) transitions. The individual hyperfine transitions are labeled (1–6 for \ce{^6Li} and 7–12 for \ce{^7Li}), and their relative frequency positions define the intrinsic spectral structure of the Li I doublet. The numerical values shown in the figure represent the frequency offsets (shifts) of each hyperfine level, and consequently determine the spacing between the allowed transitions. These separations are comparable to or smaller than broadened linewidths under many experimental conditions, leading to partial or complete overlap of individual components in measured spectra.

\section{Laser-produced plasmas}
\label{LPPs}

Optical spectroscopy performed directly on solid materials is typically unsuitable for isotopic analysis because the high atom density and condensed-phase environment broaden spectral lines, masking the small isotopic shifts described in Section~\ref{isotope_shifts}. Converting the solid into a low-density gas-phase atomic reservoir significantly reduces these broadening contributions, enabling more precise optical spectroscopy. Laser ablation can generate this atomic reservoir by focusing a high-power pulsed laser onto a solid target above the ablation threshold, generating a transient micro-plasma (a LPP), that contains neutral atoms, ions, and molecules representative of the target composition. Optical spectroscopic techniques, including emission, absorption, and fluorescence spectroscopy, can then interrogate this atomic reservoir to 
extract isotope-sensitive spectral signatures \cite{HariAPR2018}. Relative to alternative atomization approaches such as thermal or electron-beam evaporation and magnetron sputtering, LA offers rapid, minimally destructive sampling with no requirement for solution preparation, making it well suited for direct analysis of solid materials in complex or hazardous environments, or in remote/standoff configurations. The following subsections provide a concise overview of LPP generation, the key factors influencing its properties, and the associated molecular evolution, along with a summary of the spectroscopic tools used to study atomic and molecular species within the LPP reservoir. 

\subsection{Laser-produced plasma generation and key parameters}

The physics of LPP formation is complex and extensively described elsewhere \cite{Radziemski1989Book, HarilalRMP2022, miloshevsky2022ultrafast}. Briefly, laser-target interaction above the ablation threshold initiates a sequence of processes including surface heating, melting, vaporization, particle ejection, plasma formation, and rapid expansion into the 
surrounding medium. The timescales and magnitudes of these processes depend sensitively on laser parameters, including pulse duration, energy, and wavelength, as well as target properties and ambient conditions. The initial temperature and electron density of LPPs typically exceed 10,000~K and $\approx 10^{19}$~cm$^{-3}$ \cite{HarilalRMP2022, ZhangSCAB2014}, 
respectively, though both of these parameters decay rapidly as the plasma expands and cools. The mass ablated per laser shot is typically in the range of nano- to pico-grams, making LA effectively minimally destructive for most solid 
samples.

Laser pulse duration has a particularly significant effect on LPP properties relevant to isotopic analysis. Femtosecond (fs) LA generates lower-temperature plasmas with predominantly neutral atomic emissions, reduced continuum background, shorter emission persistence, and narrower angular distributions compared to nanosecond (ns) LA \cite{HariAPR2018, 2020-AC-Liz, 2014-LIBSbook-Hari}. These properties are advantageous for isotopic analysis because reduced continuum emission and lower electron densities translate directly to narrower spectral linewidths and improved signal-to-background ratios. The laser wavelength and intensity further modulate laser-target and laser-plasma coupling, influencing the ablation threshold, ionization ratio, and plasma expansion dynamics \cite{HarilalRMP2022, HariAPR2018}.

The nature and pressure of the ambient medium surrounding the LPP play a particularly important role in isotopic analysis of H and Li. The ambient gas controls the cooling rate, recombination dynamics, and degree of plasma confinement \cite{KautzJAAS2021, boulmer1993plasma, KautzOE2021}. In vacuum, the LPP expands adiabatically; in the presence of a background gas, the plume undergoes confinement, splitting, and shock formation that alter its spatial and temporal evolution \cite{2003-JAP-Hari}. Reduced ambient pressure generally produces lower electron densities at a given time delay, which directly reduces Stark broadening (a critical advantage for resolving the H Balmer series transitions, as discussed further in Section~\ref{line_broadening}). The choice of background gas is also important for controlling plasma physical conditions. For example, reduced emission linewidths were observed in He (in comparison to Ar) at comparable pressures due to more efficient cooling in He and lower Stark broadening contributions \cite{KautzOE2021}. When the LPP expands into a reactive gas such as air, plasma chemistry can drive additional changes in plume composition \cite{SkrodzkiPoP2019}, which can affect the formation of molecular species, which can be relevant to isotopic analysis in reactive material systems \cite{kautz2022oxidation}.

\subsection{Molecular species in laser-produced plasmas}

Initially, LA plumes consist primarily of excited atoms and ions. As the plasma cools below $\approx$8000~K, molecular species form through recombination and reactions with ambient species \cite{HariJAASPu, kautz2021optical, HariACS2016, alessandro2017laser, serrano2016molecular}. The formation of molecular species is governed by factors including the local atom density, ambient pressure, plasma temperature, and the thermodynamic stability of the relevant molecules \cite{kautz2022oxidation}. Some species can also be ejected directly from the target during ablation (e.g., C$_2$) \cite{phillips2025ground, iida1994optical}. For isotopic analysis, molecular emission pathways can 
complement atomic transitions. Rotational and vibrational transitions in diatomic molecules can exhibit isotopic shifts substantially larger than their atomic counterparts, making molecular spectroscopy attractive in certain cases. Molecules with lighter elements have large isotope shifts with well-separated rotational lines and vibrational bands, enabling isotopically-resolved measurements in species such as BO \cite{niki1998measurement}. Heavier diatomic molecules such as UO have many degrees of freedom and closely-spaced rotational lines, producing highly congested spectra, making isotopic shift extraction more challenging \cite{HariAPR2018}. 

\subsection{Spectroscopy methods}

\begin{figure}
    \centering
    \includegraphics[width=0.95\linewidth]{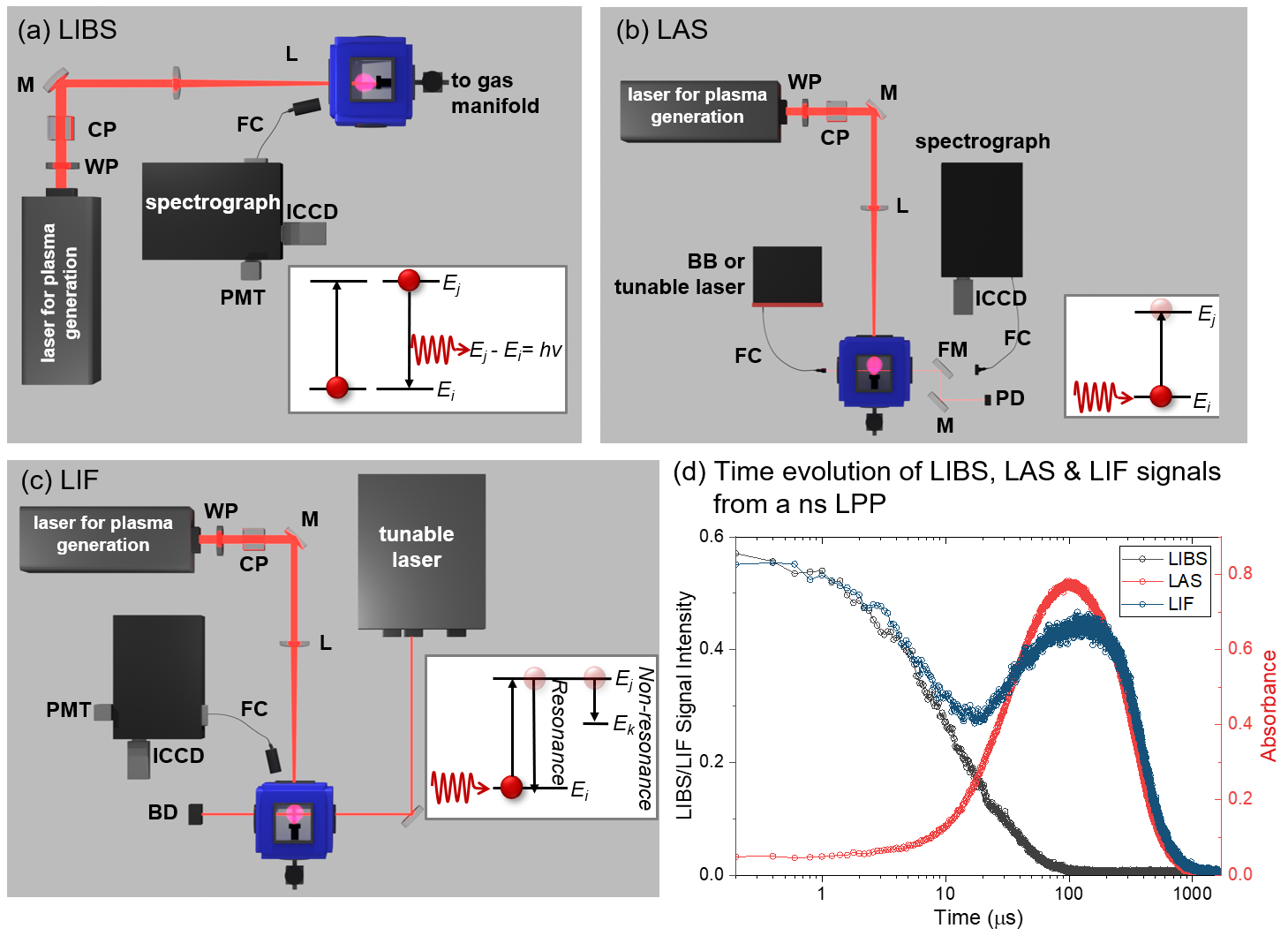}\caption{\label{fig:experimental_schematics} Overview of experimental set-ups and fundamentals of LA for plasma generation, including: (a) optical emission spectroscopy (i.e., LIBS), (b) LAS, (c) LIF, and (d) example of the time-evolution of LIBS, LAS, and LIF signals in a ns LPP. In (a), the experimental schematic shows a laser for LA/plasma generation, optics for laser energy control, focusing, and light collection, and a spectrograph coupled to an ICCD (or CCD) or PMT for light detection. In (b) an experimental set-up describes LAS, including a laser for plasma generation, optics, a second laser and optics that can either be tunable or broadband, the output from which is passed through the plasma and sent to a photodiode or a spectrograph. (c) An experimental schematic for LIF that incorporates aspects from emission and absorption spectroscopy modalities in (a) and (b). The LIF schematic includes a tunable laser source and a spectrograph that can be operated either using the ICCD or the PMT. Figure insets in (a) - (c) show schematics of electronic energy levels for LIBS, LAS, and LIF, respectively. The energy levels shown for LIF include examples of resonance or non-resonance LIF in which a directly coupled transition is used. (d) Time evolution of LIBS, LAS, and LIF signals for Al neutral atoms in a nanosecond LPP generated in 40 Torr Ar. For the LAS measurements, the probe laser was tuned to the Al I 394.40 nm transition ($0 \rightarrow 25{,}348~\mathrm{cm^{-1}}$). Because this transition shares the same upper energy level with the Al I 396.15 nm ($112 \rightarrow 25{,}348~\mathrm{cm^{-1}}$), the LIF measurements were performed by exciting the Al I 394.40 nm and detecting the directly coupled LIF emission at Al I 396.15 nm to minimize background from scattered probe laser light. LIBS signal was collected at 394.40 nm. Acronyms/abbreviations used in the figure are defined as follows: WP (wave plate), CP (cube polarizer), M (mirror), L (lens), FC (fiber cable), ICCD (intensified charge-coupled device), PMT (photomultiplier tube), LPP (laser-produced plasma), BB (broadband), FM (folding mirror), BD (beam dump).}
\end{figure}

LPPs serve as the atomic reservoir in three primary LA-hyphenated optical spectroscopic modalities: laser induced breakdown spectroscopy (LIBS) \cite{SinghLIBSbook}, laser ablation laser absorption spectroscopy (LA-LAS) \cite{MERTEN2022review, HariAPR2018}, and laser 
ablation laser induced fluorescence (LA-LIF) \cite{HariAPR2018}. Schematics of the experimental configurations for each approach are shown in Figure \ref{fig:experimental_schematics}(a), (b), and (c) for LIBS, LAS, and LIF, respectively. An example of the time evolution of LIBS, LIF, and LAS signals are reported in Figure \ref{fig:experimental_schematics}(d) for a nanosecond LPP. This example is provided for a resonance transition for Al neutral atoms in 40 Torr Ar. It should also be noted that many experimental parameters (e.g., ambient gas pressures, incident laser energy and wavelength, and pulse width) can impact the observed trends and temporal scale.

In LIBS, spontaneous emission from excited plasma species is collected and spectrally resolved, providing broadband, multi-element detection with minimal experimental complexity. In LA-LAS, a secondary light source, either broadband or tunable laser, is transmitted through the plasma and detected on the far side, measuring absorption from states with low-energies populated at delayed times when the plasma has cooled and linewidths have narrowed. In LA-LIF, a tunable laser resonantly excites a specific transition and the resulting fluorescence is collected, offering high spectral selectivity with reduced alignment constraints relative to transmission-based absorption configurations. In resonant LIF, the excitation and detected emission wavelengths are the same and care must be taken to reduce detection of scattered probe light.  In non-resonant LIF, the excitation and detected emission wavelengths are different \cite{HariPSST2021}.

Regardless of the spectroscopic modality, the spatial and temporal evolution of LPP properties (e.g., temperature, electron density) and composition strongly influences the achievable isotopic resolution. As shown in Section~\ref{line_broadening}, Stark and Doppler broadening are the primary linewidth contributions for H and Li transitions in LPPs, and both depend on the plasma conditions at the time of interrogation. For absorption spectroscopy, pressure (van der Waals) broadening may also be important.  Careful selection of the probe time window, spatial collection region, ambient pressure, and laser parameters is therefore essential for optimizing isotopic discrimination in all three spectroscopic modalities. The following sections survey the key line broadening mechanisms seen in H and Li transitions and how the LA experimental parameters have been exploited in practice for minimizing line broadening for H (Section \ref{H_isotopic_analysis}) and Li (Section \ref{Li_isotopic_analysis}) isotopic analysis.

\section{Line broadening of H I and Li I transitions}
\label{line_broadening}

A critical parameter for utilizing optical spectroscopy in isotopic analysis is the ratio of linewidth to the isotopic shift. Although LPPs as an atomic reservoir come with several advantages for isotopic analysis, the physical conditions of LPPs (high temperature and high electron density) can result in the broadening of atomic and molecular transitions, thereby complicating the measurement of small isotopic shifts. All transitions in LPPs may experience various broadening due to natural, Stark, Doppler, and pressure effects. Additionally, instrumental broadening will be present and prominent when a spectrograph system is used for light detection (e.g., in LIBS or broadband - absorption spectroscopy (BB-AS)). 

Radiative transport of light in the LPP leads to modified emission intensities and lineshapes due to effects commonly called self-absorption and self-reversal \cite{palleschi2022avoiding}. Self-absorption in homogeneous media with high peak absorptions (alternately termed high optical depths or high optical densities) leads to a nonlinear dependence of emission intensity versus atomic concentration. High absorptions most commonly occur near the line center of transitions originating from the ground state (resonance lines) but may occur for any transition with sufficiently high oscillator strength and lower level population. The combined effects of emission and absorption under these conditions can lead to self-absorption, which can modify the line profile and increase the apparent linewidth \cite{PalleschiBookChapter2025}. For spatially inhomogeneous media including LPPs, self-reversal of emission lines may occur when radiation emitted from a high-temperature plasma core passes through lower-temperature peripheral regions of the LPP before reaching the detection system. The dominant effect for self-reversal is a characteristic central dip in the line profile. It should be cautioned that while significant confusion on the origin and terminology of self-absorption and self-reversal exists \cite{palleschi2022avoiding}, their effects on intensities and lineshapes of emission lines in turn influence analytical performance, and so cannot be ignored \cite{UrbinaSCAB2022, KautzOE2023, TouchetSCAB2020, palleschi2022avoiding, soumyashree2026simulation}.

Among the various broadening mechanisms possible, the natural linewidth of H and Li atomic transitions are very small and can be assumed to be negligible. However, Stark and Doppler broadening of H and Li atomic transitions are particularly detrimental to isotopic analysis. Stark broadening, a form of collisional broadening, originates from the local electric fields generated by ions and electrons. The contribution to broadening caused by ions is relatively minor compared to the electronic component and is often not considered in analyses. The Stark width or shift depends quadratically on the electric field for most atoms and ions, but H is an exception, exhibiting a linear dependence. Hence, Stark broadening contribution is relatively high for H atomic transitions even at lower electron densities ($\approx 1 \times 10^{15} cm^{-3}$) compared to optical transitions of other elements. 

Given that LPPs can achieve electron densities exceeding 10$^{16}$ cm$^{-3}$ during the early stages of plasma evolution, substantial Stark broadening is seen in H$_\alpha$. The empirical relationship between the FWHM ($\lambda_{H_\alpha}$) of the $^{1}$H$_\alpha$ transition and electron density ($n_e$) is described as follows \cite{parigger2018laboratory}:

\begin{equation}
    \lambda_{H_\alpha}(nm) \approx  1.3\left(\dfrac{n_e(cm^{-3})}{10^{17}}\right)^{0.64 \pm 0.03}
    \label{eq:H_Stark_width}
\end{equation}

For the Li I transition at $\approx$670.8 nm, the Stark width is related to  n$_e$ through the relation \cite{Kunze2009}: 

\begin{equation}
   \lambda_{1/2}(nm)   \approx  2W \left(\dfrac{n_e (cm^{-3})}{10^{16}}\right) 
   \label{eq:Li_Stark_width}
\end{equation}
where W is the impact parameter. Impact parameters for Li 670.8 nm are 9.91 $\times$ $10^{-4}$ and 1.38 $\times$ $10^{-3}$ nm for temperatures of 5,000 and 10,000 K, respectively \cite{GriemBook2012}. The Stark broadening due to electron contribution has a Lorentzian lineshape. 

For both H and Li, Doppler broadening is significant due to the low atomic masses. Doppler broadening is caused by the random thermal motion of atoms. In a gaseous ensemble of particles with a random Boltzmann distribution of velocities, the resulting broadening is inhomogeneous and has a Gaussian lineshape with FWHM defined by \cite{demtroder2015laser}:
\begin{equation}
    \lambda_D (nm) = 7.16 \times 10^{-7}\lambda_0 (nm)\left(\dfrac{T (K)}{m (amu)}\right)^{1/2}
    \label{eq:doppler_broadening}
\end{equation}
where T is temperature in K and m is the atomic mass in amu. Since the Doppler broadening has $(T/m)^{0.5}$ dependence, it is significant for lighter elements like H and Li.  

\begin{figure}[h]
\centering
\includegraphics[width=1\linewidth]{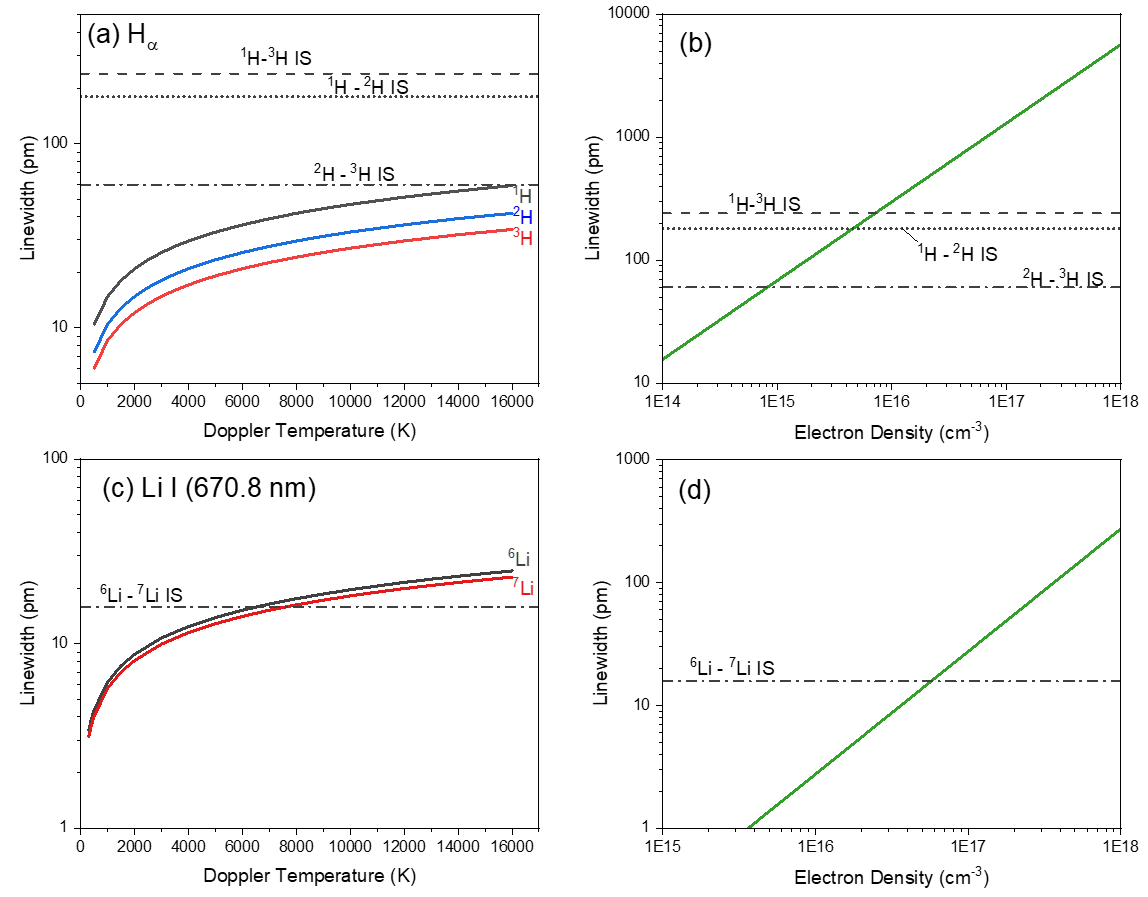}
\caption{\label{fig:line_broadening}Estimated linewidths contributed by Doppler and Stark broadening effects on (a, b) H$_{\alpha}$ and (c, d) Li I (670.8 nm) transitions at various temperatures and electron densities in the plasma. Dashed lines in each sub-figure indicate isotopic shifts (IS) of H$_\alpha$ and Li I 670.8 nm transitions. Linewidths in (a) and (c) were calculated using Equation \ref{eq:doppler_broadening}, linewidths given in (b) were calculated using Equation \ref{eq:H_Stark_width}, and linewidths reported in (d) were calculated using Equation \ref{eq:Li_Stark_width}, using the impact parameter (W) of 1.38 $\times$ $10^{-3}$ nm for a temperature of 10,000 K \cite{GriemBook2012}.} 
\end{figure}

Figure~\ref{fig:line_broadening} shows the estimated Stark and Doppler linewidths (FWHM) as a function of Doppler temperature and electron density for: H$_\alpha$ (a and b), and the Li I 670.8~nm (c and d). The solid diagonal line in panels (b and d) represents the Stark broadening contribution to linewidth as a function of electron density, while the solid curves in (a) and (c) show the Doppler broadening contribution to linewidth over a range of temperatures. For H$_\alpha$, \ce{^1H}, \ce{^2H}, and \ce{^3H} are indicated, providing a direct reference for the linewidth values that must be achieved to resolve isotopic shifts. For Li I 670.8~nm, the positions of the \ce{^6Li} and \ce{^7Li} components are similarly indicated, and illustrate the more stringent linewidth requirements for isotopic discrimination for Li.

Figure \ref{fig:line_broadening} highlights several key trends relevant to H and Li isotopic analysis. First, Stark broadening dominates for H$_\alpha$ at electron densities above $\approx 10^{16}$~cm$^{-3}$, where linewidths exceed the \ce{^1H}/\ce{^2H} isotopic shift of $\approx$180~pm, rendering isotopic discrimination impossible without significant reduction in electron density. At an electron density of $1 \times 10^{16}$~cm$^{-3}$, the Stark width of H$_\alpha$ is $\approx$298 pm, and even when electron density is reduced to $5 \times 10^{15}$~cm$^{-3}$, the Stark width of H$_\alpha$ is still $\approx$188~pm, which is comparable to the \ce{^1H}/\ce{^2H} isotopic shift. These estimated Stark width values and trends reported in Figure \ref{fig:line_broadening}(b) for $H_\alpha$ demonstrate the need for delayed gated detection or reduced-pressure conditions during LPP generation to access lower electron densities where Stark broadening is minimized. Second, for Li I 670.8~nm, Stark broadening is substantially smaller ($\approx$2.8~pm at $1 \times 10^{16}$~cm$^{-3}$, and $\approx$1.4~pm at $5 \times 10^{15}$~cm$^{-3}$, calculated using the impact parameter for a temperature of 10,000 K), but Doppler broadening becomes the dominant limitation. At a Doppler temperature of 5000~K, the linewidth is $\approx$13~pm for \ce{^7Li} (and $\approx$14~pm for \ce{^6Li}), which is nearly equal to the \ce{^6Li}/\ce{^7Li} isotopic shift of $\approx$15.8~pm. This finding means that even at relatively low temperatures, Doppler broadening is significant enough to make resolving the isotopic shift difficult for Li. In contrast, the Doppler width of H$_\alpha$ at 5000~K is $\approx$20~pm, which, while non-negligible, is small relative to the much larger H isotopic shifts. The relationships reported in Figure \ref{fig:line_broadening} illustrate that optimizing LPP physical conditions, for example through reduced ambient gas pressure, delayed detection, or choice of laser parameters (e.g., pulse width, energy), and the choice of appropriate measurement technique (e.g., LIBS vs. LAS) is essential for reliable isotopic analysis for both H and Li. Further, the relative importance of Stark versus Doppler broadening differs significantly between the two elements. 

Because the electron density and temperature of the LPP decrease as the plasma expands and cools, spectral linewidths similarly decrease at longer delays after ablation. Performing ablation in reduced-pressure environments decreases plasma confinement which also leads to lower electron density and temperatures \cite{2014-APA-Hari}. As Doppler and Stark linewidths decrease, other line broadening mechanisms such as van der Waals, resonance, and eventually natural broadening determine the minimum spectral linewidth that can be observed \cite{gornushkin1999line}.  However, emission intensities also decrease at lower pressures and later delays after ablation, which may affect the measurement signal-to-noise ratio (SNR). For optical absorption measurements, probing at later times delays and under reduced pressures typically yields a good balance between absorption strength and linewidth \cite{2021-PRE-Hari}.  For both emission and absorption measurements, optimization of LPP conditions for isotopic analysis requires balancing reduced line broadening with sufficient SNR.

Another consideration associated with plasma expansion, particularly in reduced pressure ambient gases, is the spatial segregation of different plasma species, which can result in variations in the measured elemental or isotopic abundances. Segregation of plume species is a known challenge for analytical methods that use LA for sample introduction or plasma generation \cite{ZhangJAAS2016,DiwakarJAAS2013,JacksonJAAS2003,KuhnJAAS2007,KimuraJAAS2016}. Spatial segregation of elements or isotopes with different masses has also been observed using optical spectroscopy of LPPs \cite{MohanJAAS2026,LiSCAB2023}. For example, spatial isotopic fractionation of \ce{^1H} and \ce{^2H} was observed at $1\times10^{-4}$~mbar \cite{LiSCAB2023}. Such fractionation can cause the isotopic composition to vary spatially within an expanding plume and may bias the measured isotope ratio depending on the region of the plasma being probed or the light collection geometry. Therefore, while reduced ambient gas pressure can reduce line broadening and improve the resolution of isotopic shifts, possible fractionation and measurement geometry also need to be considered for quantitative isotope analysis.

\section{Hydrogen isotopic analysis}
\label{H_isotopic_analysis}

Optical diagnostics of LPPs based on atomic emission spectroscopy (LIBS) are well suited for H isotopic measurements because several atomic transitions (given in Table \ref{tab:H_atomic_isotope_shifts}) have relatively large isotopic shifts. In particular, the H$_\alpha$ transition near 656~nm provides an isotopic shift of approximately 180~pm between \ce{^1H_\alpha} and \ce{^2H_\alpha}, and approximately 240~pm between \ce{^1H_\alpha} and \ce{^3H_\alpha}. These shifts are sufficiently large that they can often be resolved using laboratory spectrographs with moderate resolving power, provided that line broadening is minimized. In contrast, optical absorption spectroscopy of atomic H is challenging because the ground state transition lies in the VUV spectral region, and the detectability of other transitions depends on whether the corresponding energy levels are sufficiently populated. Molecular emission spectroscopy from LPPs has been used to detect H isotope shifts between \ce{O^1H} and \ce{O^2H} using the Laser Ablation Molecular Isotope Spectroscopy  (LAMIS) method \cite{RussoSCAB2011}.  While molecular absorption spectroscopy has not to our knowledge been used to detect H isotopes in LPPs, there are many reports of H isotopologue detection in a wide range of gas-phase species containing \ce{^1H} and \ce{^2H}. Of particular interest is detection of tritiated water vapor \ce{^3H_2O} using infrared laser spectroscopy methods. Raman spectroscopy has also been used for detection of H isotopologues in molecules. In the next sections, we provide an overview of these methods and example results. It should be noted that in contrast to radiation detection methods, optical spectroscopy can use \ce{^2H} as a spectroscopic and chemical surrogate for \ce{^3H}. The use of \ce{^2H} as a surrogate assumes that differences in electronic excitation and ionization behavior between \ce{^2H} and \ce{^3H} do not substantially affect the optical measurement. However, the difference in isotope mass can influence physical processes during LA and subsequent plume expansion, including velocities of plume species.

\begin{table*} 
\caption{Summary of literature reporting analysis of H isotopes (\ce{^1H}, \ce{^2H}, \ce{^3H}) via optical emission spectroscopy of LPPs (i.e., LIBS) in the UV-VIS spectral region, including information on target composition, isotopes detected or analyzed, experimental details and remarks. All studies were performed using the H$_\alpha$ transition and conventional single-pulse (SP) LIBS, unless otherwise noted. SP/DP-LIBS denotes studies employing both single- and double-pulse LIBS configurations. Acronyms used in the table are defined as follows: LOD (limit of detection), SNR (signal-to-noise ratio), CF-LIBS (calibration-free LIBS), ML (machine learning), TEA (transversely excited atmospheric), LA (laser ablation), MIP (microwave-induced plasma), DPSS (diode-pumped solid-state), MW-LIBS (microwave-assisted LIBS), and PLS (partial least squares).}
\label{tab:H_Lit_Review}
\centering
\renewcommand{\arraystretch}{1.18}
\setlength{\tabcolsep}{4pt}

\begin{tabular}{p{2.0cm} p{1.1cm} p{8cm} p{1.5cm}}
\hline
\textbf{Target} &
\textbf{Isotopes} &
\textbf{Experiment details / remarks} &
\textbf{Ref.} \\
\hline

\multirow{12}{*}{Zircaloy-4}
& \multirow{7}{*}{\ce{^{1,2}H}}
& He gas (5--760 Torr), Nd:YAG (355 nm, 1064 nm, 0.02--8 ns FWHM, 2--75 mJ), SP/DP-LIBS; \ce{^1H}/\ce{^2H} resolution improved in low pressure He, DP enhances \ce{^2H} signal, \ce{^2H} LOD $\approx$10--20 ppm
& \cite{KurniawanAC2006, PardedeAC2018, KurniwanAC2012, PardedeSciRep2021, HedwigJAP2010, MarpaungJAP2011} \\
\cline{3-4}

&
& Ar, He gas (0.1--100 Torr), Nd:YAG (1064 nm, 6 ns, 10--400 J/cm$^2$), Ti:Sapphire (800 nm, 35 fs, 0.3--60 J/cm$^2$), SP/DP-LIBS; fs LPP in He minimizes line broadening, \ce{^1H}/\ce{^2H} emission localized near target, DP enhances \ce{^2H} signal $\approx$10x $\&$ reduces self-absorption
& \cite{ KautzOE2021, KautzJAAS2021, kautz2021interplay, KautzSCAB2024, ShaikPoP2024, HarilalSPIE2024} \\
\cline{2-4}

& \multirow{3}{*}{\ce{^{1,2,3}H}}
& Ar gas (26--180 Torr), Ti:Sapphire
(800 nm, 35 fs, $\approx$160 J/cm$^2$); detection of \ce{^3H} using ultrafast LIBS; depth profiling revealed non-uniform \ce{^3H} distribution
& \cite{HarilalJAAS2024} \\
\hline

\multirow{5}{*}{W}
& \multirow{5}{*}{\ce{^{1,2}H}}
& Ar and air (atmospheric pressure), fs laser (343 nm, 500 fs, 31 J/cm$^2$); CF-LIBS quantified \ce{^2H} ($\approx$1.7 at.\%);
depth resolution $\approx$600 nm/pulse
& \cite{MittelmannSciRep2023} \\
\cline{3-4}

&
& Ar, He, air (atmospheric pressure), Nd:YAG (1064 nm, 8 ns, $\approx$52 J/cm$^2$); Ar provides higher SNR than He $\&$ air; simulated \ce{^2H}/\ce{^3H} peaks through spectral fitting
& \cite{AlmavivaJNE2025} \\
\hline

\multirow{1}{*}{W--Al}
& \multirow{1}{*}{\ce{^{1,2}H}}
& Ar (3--750 Torr), Nd:YAG
(532 nm, 8 ns, $\approx$7 J/cm$^2$)
& \cite{ParisPS2017} \\
\hline

\multirow{2}{*}{Be--W, Al--W}
& \multirow{2}{*}{\ce{^{1,2}H}}
& Vacuum ($\approx$7.5$\times10^{-7}$ Torr)/Ar (0.375 Torr), Nd:YAG
(1064 nm, 5 ns, $\approx$4.3--4.5 J/cm$^2$)
& \cite{SuchovnovaNME2017, VeisPS2020} \\
\hline

\multirow{2}{*}{W--Be}
& \multirow{2}{*}{\ce{^{1,2}H}}
& Synthetic LIBS spectra (200--700 nm), ML; H-retention quantification
& \cite{GasiorSCAB2023} \\
\hline

\multirow{4}{*}{Ti}
& \multirow{2}{*}{\ce{^{1}H}}
& Vacuum, TEA CO$_2$ laser ($\sim$50 MW/cm$^2$); 
\ce{^1H} detected in Ti with native $\&$ loaded \ce{^1H}
& \cite{TrticaJAS2024} \\
\cline{2-4}

& \multirow{2}{*}{\ce{^{1,2}H}}
& He gas (22.5 Torr), Nd:YAG (1064 nm, 10 ns, $\sim$1144 J/cm$^2$); analysis of \ce{^1H}/\ce{^2H} using CF-LIBS
& \cite{XingNME2022} \\
\hline

C + \ce{^2H} on Si
& \ce{^{1,2}H}
& Ar ($\approx$11 Torr), LA--MIP, Nd:YAG
(1064 nm, 6 ns, 5.4 J/cm$^2$); predicted \ce{^2H}/\ce{^3H} discrimination
& \cite{VujadinovicEPhys2026} \\
\hline

\ce{^2H_2O}-doped graphite/\ce{SiO2}
& \ce{^{1,2}H}
& Ar and He gas ($\approx$2--60 Torr), TEA \ce{CO2}
(10.6 $\mu$m, 80 ns, 260--420 mJ)
& \cite{TraparicSCAB2024} \\
\hline

\multirow{3}{*}{\ce{H2} gas}
& \multirow{3}{*}{\ce{^{1,2}H}}

& Nd:YAG (1064 nm, 10 ns, $\approx$300 mJ)
& \cite{DUlivoSCAB2006} \\
\cline{3-4} 

&
& Ar (130--240 kPa), LIBS, Nd:YAG (1064 nm, 9 ns, 50--150 mJ);
\ce{^1H}/\ce{^2H} LODs $\approx$ 12 ppm
& \cite{AndrewsJACS2024, KitzhaberAppliedSpec2026} \\
\hline

\multirow{10}{*}{\ce{H_2O}}
& \multirow{10}{*}{\ce{^{1,2}H}}
& Air (atmospheric pressure), Nd:YAG
(1064 nm, 10 ns, $\sim$50 J/cm$^2$);
PLS enabled quantitative \ce{^2H}/\ce{^1H} analysis
& \cite{DoucetJAAS2011} \\
\cline{3-4}

&
& Air (atmospheric pressure), DPSS laser
(1064 nm, 7--9 ns, 14 mJ);
\ce{^1H}/\ce{^2H} resolved using a compact field-deployable instrument
& \cite{CremersAS2012} \\
\cline{3-4}

&
& Air (atmospheric pressure), Nd:YAG (532 nm, 6 ns, 10 mJ), MW-LIBS;
\ce{^2H} LOD $\approx$ 0.4 vol. \%
& \cite{AlamriJAAS2025} \\
\cline{3-4}

&
& He ($\approx$1 Torr), SP-LIBS, TEA CO$_2$ laser (10.6 $\mu$m, 200 ns, 500 mJ)& \cite{IdrisJJAP2004} \\
\cline{3-4}

&
& He gas (760 Torr), Nd:YAG
(1064 nm, 8 ns, 120 mJ)
& \cite{KurniawanJAP2005} \\
\cline{3-4}

&
& Air (760 Torr), Nd:YAG
(1064 nm, 10 ns, 55--75 mJ), SP/DP-LIBS (colinear); DP reduces line broadening
& \cite{BurgerPoP2018} \\
\cline{3-4}

&
& $7.5\times10^{-5}$-- 0.075 Torr, Nd:YAG
(1064 nm, 8 ns, $\approx$6 J/cm$^2$); separation of \ce{^1H}/\ce{^2H} in low pressure
& \cite{LiSCAB2023} \\
\hline

\end{tabular}
\end{table*}


\subsection{Laser Induced Breakdown Spectroscopy}

Table~\ref{tab:H_Lit_Review} summarizes representative studies reporting LIBS spectroscopy of H isotopes in LPPs. The table includes information on the analyzed matrix, observed spectral features, and selected experimental conditions. Several trends emerge from this compilation, including: (1) most studies rely on the Balmer-$\alpha$ transition because its relatively large isotopic shift enables discrimination between \ce{^1H} and \ce{^2H} using moderate-resolution spectrographs, (2) many experiments have focused on materials relevant to nuclear fusion and fission energy materials (e.g., tungsten, beryllium, and Zr alloys), highlighting the importance of H isotope monitoring for its retention and hydride formation, and (3) reported studies span a wide range of plasma excitation schemes and ambient conditions, illustrating the flexibility of optical spectroscopy for H isotope detection and analysis.

Despite this advantage of relatively large isotopic shifts for several H atomic and molecular transitions, challenges remain that complicate H isotope measurements using LIBS. First, significant line broadening occurs in LPPs, particularly due to Stark broadening arising from high electron densities. As discussed in Section \ref{line_broadening}, Stark broadening of the H$_\alpha$ transition can exceed Doppler broadening under atmospheric pressure conditions, often masking the isotopic shift. Second, the H$_\alpha$ transition originates from a high upper energy level (\(\approx 12.09\)~eV), which is populated during the early stages of plasma evolution when temperatures and electron densities are highest. Consequently, measurements performed shortly after plasma onset typically exhibit broad spectral profiles that can obscure isotopic splitting \cite{KautzOE2021, ShaikPoP2024}. Third, contamination from ubiquitous environmental protium can also impact quantitative isotopic analysis \cite{DoucetJAAS2011, CremersAS2012}. Lastly, analytical capabilities for H isotopes via LIBS can be limited. The limit of detection for H isotopes has been reported on the order of ppm \cite{PardedeAC2018}. At higher analyte concentrations, self-absorption may distort spectral lineshapes and reduce quantitative accuracy \cite{KautzSCAB2024}.

\begin{figure}
    \centering
    \includegraphics[width=0.9\linewidth]{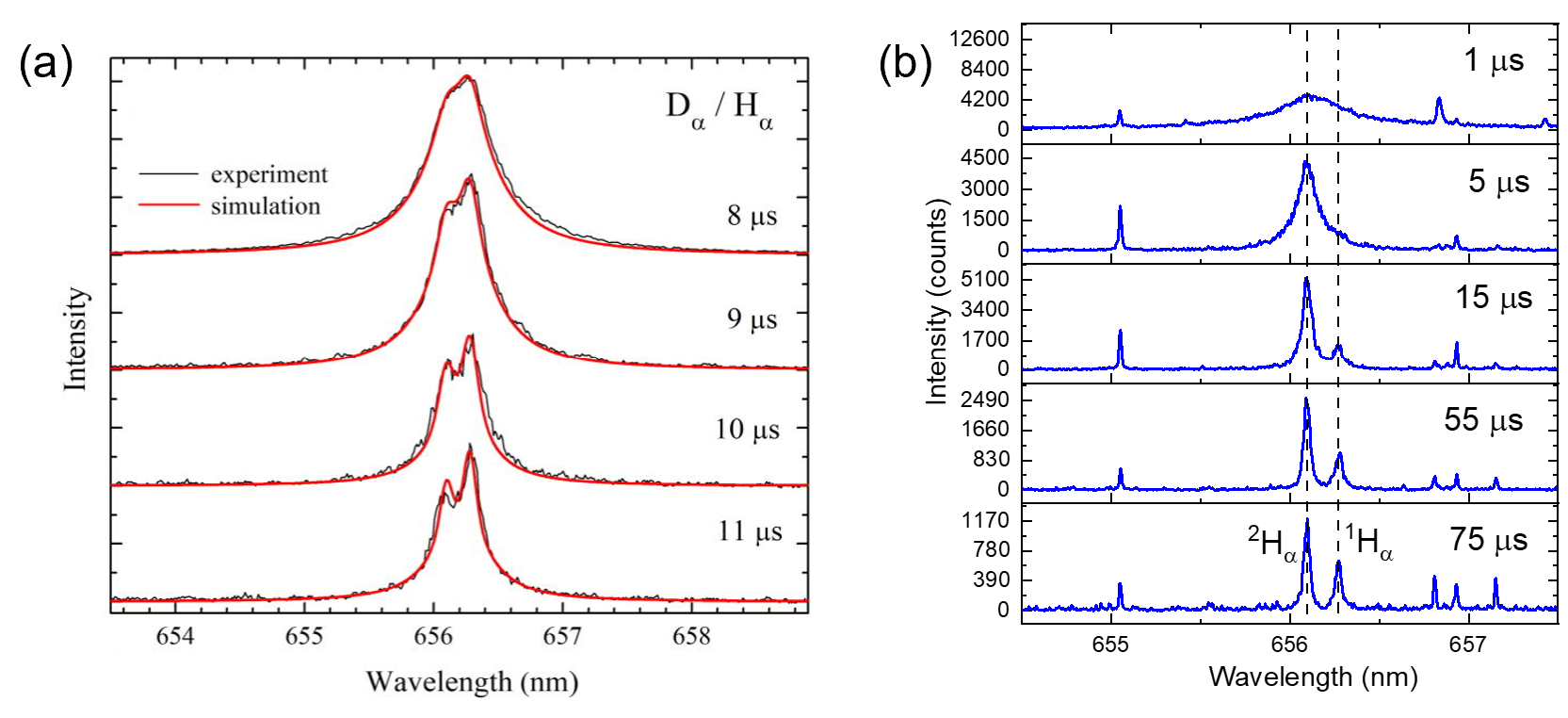}
       \caption{\label{H-TR} Time-resolved emission spectra of the Balmer-\(\alpha\) lines for protium (\ce{^1H_\alpha}) and deuterium (\ce{^2H_\alpha}) following LPP onset. (a) Experimental and simulated line profiles from a mixed \ce{^1H}/\ce{^2H} plasma, showing progressive line narrowing between 8--11~\(\mu\)s after plasma initiation using a gate width of 1 $\mu$s \cite{BurgerPoP2018}. Spectra given in (a) were generated via LA of a frozen \ce{^1H_2O}/\ce{^2H_2O} mixture containing 40 at.\% \ce{^2H} in atmospheric pressure ($\approx$750 Torr) air. (b) Emission spectra from a Zircaloy-4 target charged with \ce{^2H} gas (containing $\approx$ 10,000~ppm by mass \ce{^2H}). The gate delays are given in each sub-figure; gate widths are 0.5~\(\mu\)s for 1~\(\mu\)s delay, 1~\(\mu\)s for 5 and 15~\(\mu\)s, and 5~\(\mu\)s for 55 and 75~\(\mu\)s \cite{ShaikPoP2024}.}
   \end{figure}

Time-resolved (i.e., gated) detection and operation under reduced-pressure conditions can mitigate some of the above mentioned constraints. As the plasma expands and cools, electron density decreases and linewidths narrow, allowing partial or complete separation of \ce{^1H_\alpha} and \ce{^2H_\alpha}, for example (or \ce{^1H_\alpha} and \ce{^3H_\alpha} or \ce{^2H_\alpha} and \ce{^3H_\alpha}). However, this improvement in reduced line broadening (leading to improved isotopic discrimination) comes at the cost of reduced signal-to-noise ratio due to lower emission intensities. Figure \ref{H-TR} illustrates the time-resolved evolution of the \ce{^1H_\alpha} and \ce{^2H_\alpha} emission following LPP formation in (a) atmospheric pressure air \cite{BurgerPoP2018} and (b) flowing Ar at reduced pressure (30 Torr) \cite{ShaikPoP2024}. In both cases and at early times, Stark broadening dominates and the isotopic splitting is unresolved, as shown by the spectra collected at 8~\(\mu\)s and 1~\(\mu\)s after plasma onset in Figures~ \ref{H-TR}(a) and \ref{H-TR}(b), respectively. As the plasma expands and cools, the linewidths become narrower, and the isotopic splitting is partially resolved under atmospheric pressure conditions, whereas a distinct and persistent resolution of \ce{^2H_\alpha}/\ce{^1H_\alpha} is achieved in reduced pressure environments.

\begin{figure}
    \centering
    \includegraphics[width=1\linewidth]{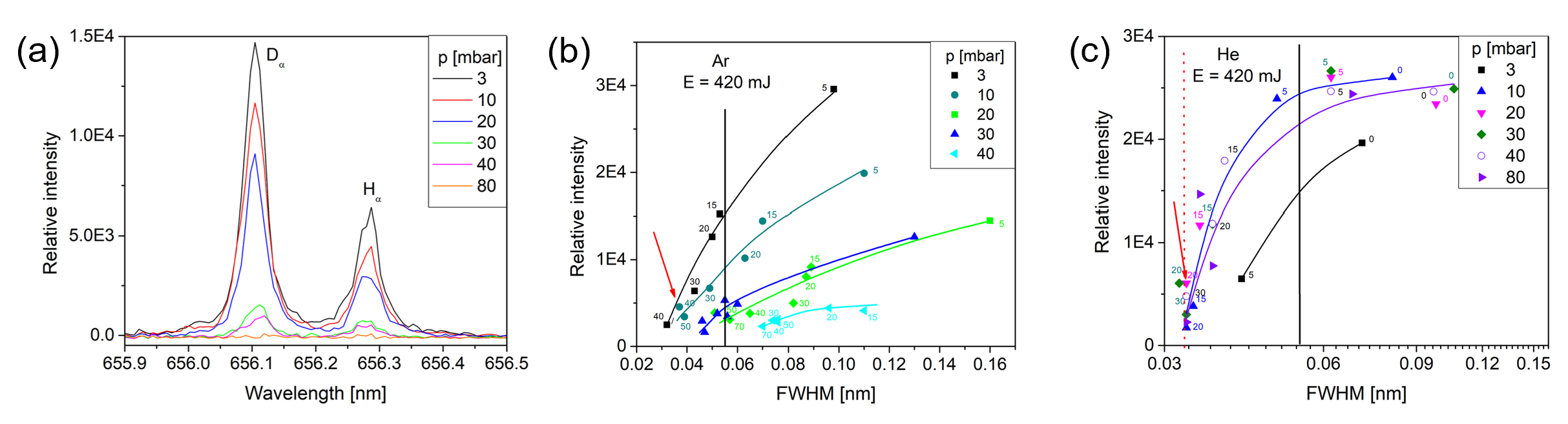}
        \caption{\label{H-linewidth-He-Ar}(a) LIBS spectra generated from a \ce{^2H}-enriched graphite target, showing the effect of He gas pressure (3–80~mbar) on lineshape at a gate delay/width of 15~\(\mu\)s / 5~\(\mu\)s. The \ce{^2H}-enriched targets were prepared from graphite and \ce{SiO2} powders (3:1 ratio) doped with \ce{^2H_2O}. LPPs were generated using a 10.6~$\mu$m, 80~ns pulse width CO$2$ laser. (b, c) Dependence of spectral line peak intensity on FWHM for different gas pressures of Ar and He, respectively. Vertical lines indicate the maximum spectral linewidth (0.054~nm) compatible with resolving \ce{^2H_\alpha} and \ce{^3H_\alpha} with equal peak intensity. Data point labels give delay times. Sub-figures adapted from \cite{TraparicSCAB2024}.}
\end{figure}

The composition and pressure of the background gas strongly influences spectral characteristics such as emission intensity and linewidth. As summarized in Table~\ref{tab:H_Lit_Review}, many studies employ reduced-pressure environments or inert background to improve isotopic discrimination. Reduced-pressure conditions generally
produce narrower spectral lines due to lower electron densities, and He backgrounds have been shown to yield narrower linewidths than Ar under comparable conditions \cite{KautzOE2021, TraparicSCAB2024}. These effects
are illustrated in Figure~\ref{H-linewidth-He-Ar}. In Figure~\ref{H-linewidth-He-Ar}(a), LIBS spectra obtained from a \ce{^2H}-enriched graphite target at He pressures ranging from 3 to 80~mbar (2.25–60~Torr) were recorded at a gate delay/width of 15~\(\mu\)s / 5~\(\mu\)s. Increasing pressure led to progressively broader linewidths but reduced emission intensity of \ce{^1H_\alpha} and \ce{^2H_\alpha}. Figures~\ref{H-linewidth-He-Ar}(b) and \ref{H-linewidth-He-Ar}(c) quantify this behavior by plotting peak intensity versus FWHM for a range of pressures and delays in Ar and He, respectively. In both gases, lower pressures yielded narrower linewidths (at comparable delay times), reflecting reduced electron density. At a given pressure, He produced consistently narrower lines than Ar, further enhancing resolvability of isotopic splitting. Overall, these studies showed that operating at reduced pressure enables clear isotopic splitting and longer persistence, improving H isotope detection by LIBS.

The target matrix is an important consideration when comparing the H isotope measurements summarized in Table~\ref{tab:H_Lit_Review}. Differences in ablation efficiency can influence plasma temperature, electron density, species number density, and plume expansion \cite{ZhangSCAB2014, OjedaAdvMatInt2018}. Although the intrinsic isotope shift for a given transition is independent of the target material, the measured linewidth and SNR can be affected by these matrix dependent plasma conditions. Therefore, ambient pressure, detection delay, and other experimental conditions optimized for isotopic resolution in one target material may not provide the same performance for another. Systematic studies comparing isotopic resolution across different target materials under similar experimental conditions are currently limited and would be useful for understanding these matrix effects.

The target material and its processing history can also influence the amount and distribution of the H isotope being measured. For example, the concentration and distribution of \ce{^2H} or \ce{^1H} in Zircaloy-4 will depend on how the isotope was introduced into the material, such as through high temperature gas loading \cite{KautzJAAS2021}, autoclave corrosion, or corrosion in light or heavy water reactors (e.g., CANDU reactors), where Zircaloy-4 is used as fuel cladding. Similarly, prior work on W targets has demonstrated that \ce{^2H} retention varies with W grade and material condition \cite{BuziJNM2015,SugiyamaPhysicaScripta2014,BaldenJNM2014}. Different methods used to introduce or retain H isotopes in W can also influence their amount and distribution in the target material. Examples include co-deposition with \ce{^2H} using High-Power Impulse Magnetron Sputtering \cite{AlmavivaJNE2025}, vacuum arc-discharge deposition \cite{AlmavivaFusionEngDes2020}, and exposure using linear plasma generators \cite{JogiJNM2021,MittelmannSciRep2023,VanDerMeidenNuclearFusion2021}. These differences in isotope concentration and distribution can influence the measured signal intensity, in addition to experimental conditions and light collection geometry, and should be considered when comparing results obtained from different samples.

Another challenge associated with LIBS is its relatively low sensitivity compared to laboratory-based mass spectrometry methods. In one study, the LIBS detection limit for deuterium (\ce{^2H}) in a solid matrix was shown to be $\approx$10~ppm \cite{PardedeAC2018}. Conversely, at higher concentrations, self-absorption can become problematic due to increased atom density of the analyte, complicating quantitative isotopic analysis \cite{KautzSCAB2024}. To address these challenges, several advanced methodologies have been explored, including microwave-enhanced LIBS \cite{ikeda2022spatially, AlamriJAAS2025}, LIBS combined with static electric fields \cite{karnadi2025highly},  and double-pulse (DP) excitation \cite{PardedeAC2018, KautzSCAB2024}. 

\begin{figure}
    \centering
    \includegraphics[width=1\linewidth]{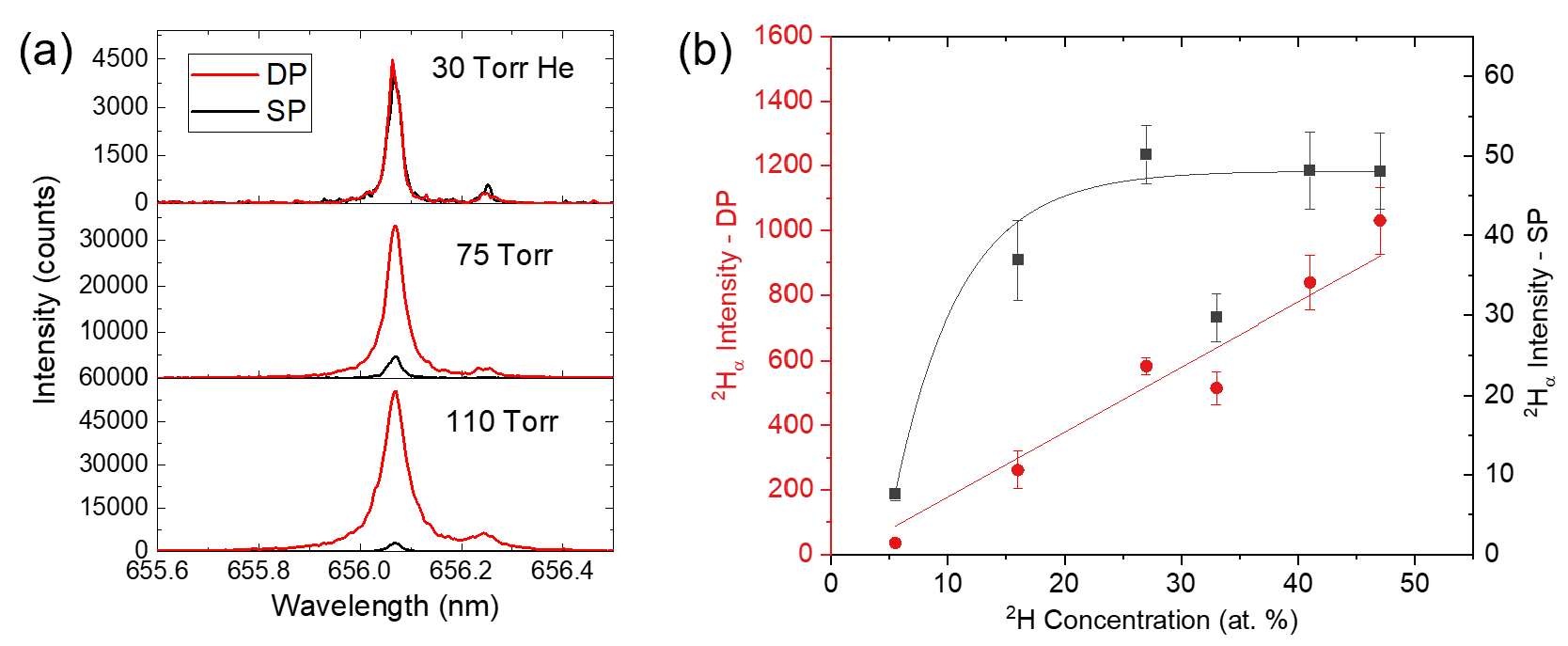}
    \caption{\label{fig:H-Isotopes-DP} (a) \ce{H_\alpha} emission spectra collected using SP and DP schemes in He background pressures of 30, 75, and 110~Torr. LPPs were generated from \ce{^2H}-charged Zircaloy-4 using a 1064~nm, 6~ns Nd:YAG laser with a fluence of $\approx$25~J/cm$^2$. All spectra were acquired using a 500~mJ heating-pulse energy, a 500~ns inter-pulse delay (IPD), and a 6~$\mu$s / 10~$\mu$s gate delay/width. For the DP configuration, the second laser pulse generated a He breakdown plasma $\approx$1.5~mm above the target to reheat the ablation plasma. (b) Calibration curves relating the \ce{^2H_\alpha} emission intensity to \ce{^2H} concentration for both SP and DP configurations, obtained under identical laser energies and timing parameters in a 75~Torr He environment. The Zircaloy-4 specimens used to generate calibration curves in (b) contained $\approx$5.5--47~at. \% \ce{^2H}, introduced by gas charging at 500~$^\circ$C, the experimental procedure for which is detailed elsewhere \cite{KautzJAAS2021}. Spectra were acquired with a spectral resolution of $\approx$13~pm at 632.8 nm \cite{KautzSCAB2024}.}
    
\end{figure}

Among the various methods used to enhance LIBS sensitivity, the DP reheating scheme offers a wide range of parameters for optimization. There are several DP schemes, including collinear, orthogonal, and crossed-beam approaches \cite{diwakar2013role}. The orthogonal reheating scheme, in particular, is effective for increasing LIBS signals and mitigating self-absorption in \ce{^2H_\alpha} transition  \cite{KautzSCAB2024, KurniawanJAP2005, FantoniSCAB2017}. By optimizing parameters such as inter-pulse delay and heating-laser energy, orthogonal DP scheme enhanced emission, minimized line broadening, and reduced self-absorption at higher analyte concentrations.  Figure~\ref{fig:H-Isotopes-DP}(a) compares \ce{^2H_\alpha} emission spectra obtained with single-pulse (SP) and DP configurations at several He background pressures, highlighting the substantial signal enhancement achieved using a DP scheme. Figure~\ref{fig:H-Isotopes-DP}(b) shows calibration curves relating \ce{^2H_\alpha} emission intensity to \ce{^2H} concentration for both SP and DP, demonstrating that the DP scheme mitigates self-absorption and extends the analytical capabilities of the LIBS approach to be applied to low and high analyte concentrations. Beyond traditional nanosecond (ns) and femtosecond (fs) LIBS, Li \textit{et al.} \cite{li2020calibration} demonstrated the effectiveness of fs laser filamentation combined with chemometric analysis to enhance peak resolution for \ce{^2H_\alpha} and \ce{^1H_\alpha} lines while minimizing spectral broadening. 

LIBS has emerged as a valuable tool for H isotope measurements in nuclear materials and fusion environments because of its non-contact operation, rapid analysis capability, multi-element detection, and potential for depth-resolved measurements \cite{WuJPD2019, Hongbin2016review, sahithya2025application, TraparicSCAB2024, ParisPS2017, MittelmannSciRep2023, nishijima2021dynamic, sun2021ex, vujadinovic2025hydrogen, TrticaJAS2024, wust-2024, RistkokNME2025, maurya2020review, VanDerMeidenNuclearFusion2021, AlmavivaJNE2025, VeisPS2020, SuchovnovaNME2017, XingNME2022}. The technique has been applied to plasma facing components, reactor wall materials, and fusion-relevant environments for assessing elemental composition, H isotope retention, deposited layer thickness, and isotope depth distributions \cite{zhao2018remote,piip2017loading,jiang2019upgraded,colao2017libs,lopez2016libs, Hongbin2016review}. Several studies have therefore demonstrated or developed LIBS systems for in situ analysis of plasma facing components in fusion devices such as the Experimental Advanced Superconducting Tokamak (EAST) \cite{zhao2018remote,LiuFusionEngrDes2017}, Joint European Torus (JET) \cite{YiNME2025,KarhunenJNM2015, RistkokNME2025}, Tungsten (W) Environment in Steady-state Tokamak (WEST) \cite{FavrePhysScripta2024}, and Tokamak Experiment for Technology Oriented Research (TEXTOR) \cite{HuberPhysScripta2011}, with relevance and potential application to the International Thermonuclear Experimental Reactor (ITER). In these configurations, the laser is delivered remotely to the plasma facing surface and the resulting LPP emission is collected through an optical collection system, allowing measurements to be performed without removing the component from the fusion device. These studies highlight the utility of LIBS for in situ monitoring and characterization of H isotope implantation, retention, and distribution in materials relevant to nuclear systems.

\begin{figure}
   \centering
  \includegraphics[width=1\linewidth]{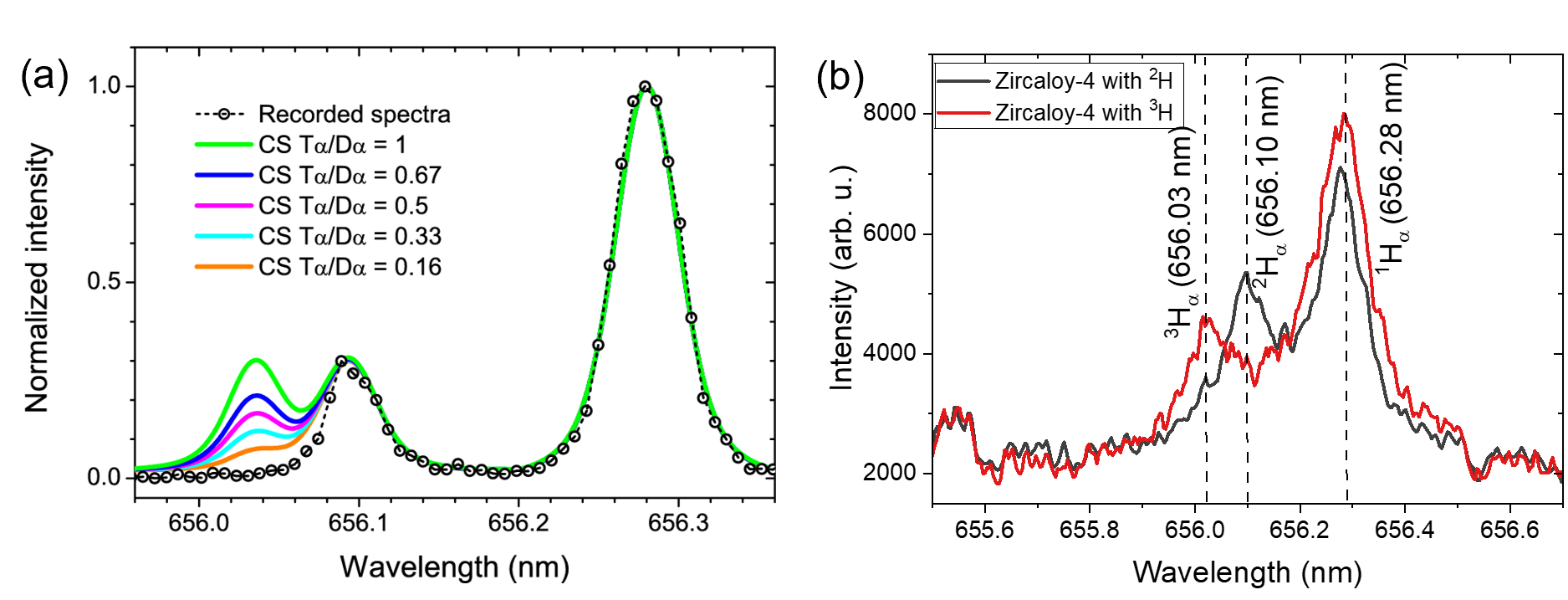}
   \caption{\label{fig:H-D-T spectrum} (a) Theoretical estimation of \ce{^3H} resolving capabilities plotted with experimentally measured spectra from a low pressure microwave-induced plasma (MIP) generated from a Si target coated with a carbon and \ce{^2H} layer. Material was introduced into the Ar MIP by LA using a 1064~nm, 6~ns Nd:YAG laser. The MIP was generated using a 2.45~GHz microwave source. Sub-figure adapted from \cite{VujadinovicEPhys2026}. (b) Experimental data from LIBS with $H_{\alpha}$ lines labeled for \ce{^1H}, \ce{^2H}, \ce{^3H}. LPPs were generated using an 800~nm, $\approx$35~fs Ti:Sapphire laser with an ablation fluence of $\approx$210~J/cm$^2$, and spectra were collected with a spectral resolution of $\approx$13~pm. The \ce{^1H_\alpha} and \ce{^2H_\alpha} lines were measured from a \ce{^2H}-loaded Zircaloy-4 target containing $\approx$4300~ppm \ce{^2H}, while the \ce{^1H_\alpha} and \ce{^3H_\alpha} lines were measured from a neutron-irradiated, Zircaloy-4 target (all experiments were performed in 45~Torr Ar). Sub-figure adapted from \cite{HarilalJAAS2024}.}
\end{figure}

Most LIBS studies of H isotopes have focused on distinguishing \ce{^1H} and \ce{^2H} due to the relative accessibility of these isotopes. Experimental studies involving \ce{^3H} remain limited because of its radiotoxicity and associated radiological constraints. Theoretical analysis by Vujadinovic \textit{et al.} \cite{VujadinovicEPhys2026} evaluated the resolving capability for \ce{^3H} by comparing calculated spectra with experimentally measured \ce{^1H_\alpha} and \ce{^2H_\alpha} emission profiles, as illustrated in Figure~\ref{fig:H-D-T spectrum}(a). Recent work by Harilal \textit{et al.} \cite{HarilalJAAS2024} demonstrated the feasibility of detecting \ce{^3H} using LIBS during depth profiling of neutron-irradiated Zircaloy-4. In that study, ultrafast laser ablation using 35~fs pulses in a 45~Torr Ar ambient produced sufficiently narrow emission profiles to resolve the isotopic splitting between \ce{^1H_\alpha}, \ce{^2H_\alpha}, and \ce{^3H_\alpha}. Representative spectra from deuterium-charged and neutron-irradiated Zircaloy-4 samples are shown in Figure~\ref{fig:H-D-T spectrum}(b). 

\begin{figure}
   \centering
  \includegraphics[width=1\linewidth]{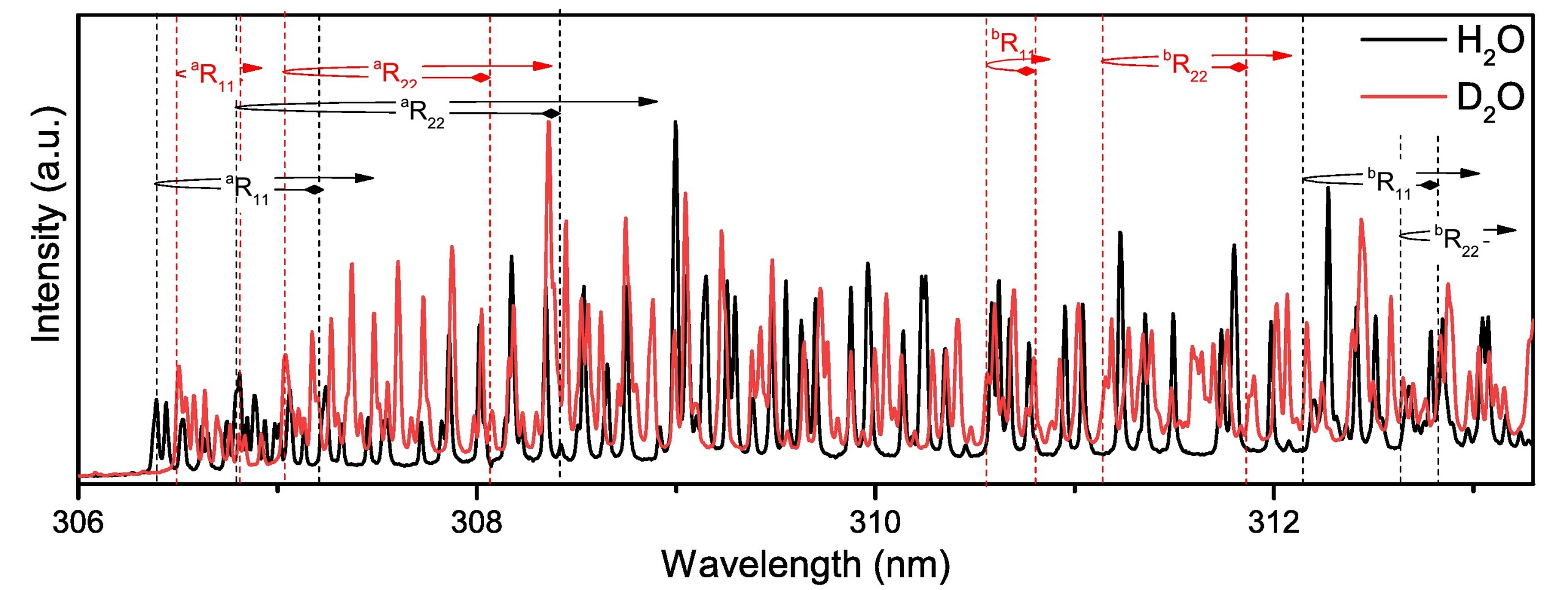}
   \caption{\label{fig:H_molecular_spectra} Representative spectra of \ce{O^1H} and \ce{O^2H} from LPPs generated from \ce{^1H_2O} and \ce{^2H_2O}. LPPs were generated using an 800~nm, $\approx$35~fs Ti:Sapphire laser with a pulse energy of 4~mJ. Light emitted by the LPP was collected using a spectrograph--ICCD with resolution of $\approx$10~pm. The spectra show the UV \ce{O^1H}/\ce{O^2H} \ce{A^2\Sigma+ -> X^2\Pi} molecular bands, where \ce{^aR} and \ce{^bR} indicate the R-branch bandhead features from the $(0,0)$ and $(1,1)$ bands, respectively. Figure reproduced from Reference~\cite{RanSCAB2021}.}
\end{figure}

While the studies summarized in Table~\ref{tab:H_Lit_Review}
focus on atomic emission, several investigations have demonstrated
that molecular emission from \ce{O^1H} and \ce{O^2H} radicals can provide an alternative route for isotopic analysis in LPPs \cite{RussoSCAB2011, RanSCAB2021, bol2016laser, BolshakovJAAS2017, BolshakovSPIE2012, ChoiSCAB2019}. The \ce{O^1H} molecule exhibits an emission band in the 306–318~nm range corresponding to the A\(^2\Sigma^+\)–X\(^2\Pi\) system. Vibrational and rotational isotopic shifts can be either positive or negative, depending on the upper and lower state quantum numbers. For example, the bandhead shifts of the R\(_{11}\) and R\(_{22}\) branches of \ce{O^2H} relative to \ce{O^1H} (shown in Figure \ref{fig:H_molecular_spectra} \cite{RanSCAB2021}) are +0.123~nm and +0.220~nm, whereas the Q\(_{11}\) and P\(_{11}\) branches exhibit shifts of –0.687~nm and –0.838~nm, respectively. Because the OH spectrum is highly congested (containing 12 branches and 6 satellite transitions) high-resolution spectrographs combined with spectral modeling are essential for accurate isotopic analysis \cite{RanSCAB2021, SarkarSCAB2013}. Use of molecular species for isotope ratios in LPPs should also consider plasma chemistry, to optimize formation of the targeted molecular species. 

\subsection{Absorption Spectroscopy}

To date, atomic absorption methods have not been reported for isotopic analysis of H from solids using LA methods. Absorption spectroscopy of H isotopes is challenging due to the electronic structure.  Transitions from the ground state of H occur at VUV wavelengths $\sim$ 121.56 nm or shorter wavelengths, and are difficult to measure with conventional absorption methods. For transitions at longer wavelengths in the visible, the higher lower state energy levels limit the applicability of absorption spectroscopy methods.  For example, the lower energy level of the H\(_\alpha\) transition (82259.105~cm\(^{-1}\), 10.2~eV) \cite{Kurucz_database} results in a very low thermally-excited population available to probe at low temperatures. However, plasma sources such as LPPs may provide sufficient excitation at high temperatures, or may support longer-lived metastable populations to probe. 


Several groups demonstrated the application of two-photon absorption laser induced fluorescence (TALIF) for detecting H isotopes \cite{bokor1981two,cesar1996two, LIF-review-Yevegeny}. Kulatilaka \textit{et al.} \cite{kulatilaka2008comparison,schmidt2016femtosecond} used TALIF to obtain spatially and temporally resolved measurements of atomic H in moderate pressure pulsed discharge plasmas. TALIF has also been applied to measure deuterium in fusion plasmas \cite{KajitaNME2025,MerkJPD2023}. Two-photon LIF approaches have also been investigated for measurements of atomic \ce{^2H} and \ce{^3H} in magnetically confined fusion plasmas \cite{VoslamberRevSciInst1998, VoslamberRevSciInst2000}. For example, Doppler-free two-photon excitation of the Lyman-$\alpha$ transitions combined with neutral beam injection has been proposed for local measurements of \ce{^1H}/\ce{^2H}/\ce{^3H} density ratios \cite{VoslamberRevSciInst1998}.

\begin{figure}
   \centering
  \includegraphics[width=0.8\linewidth]{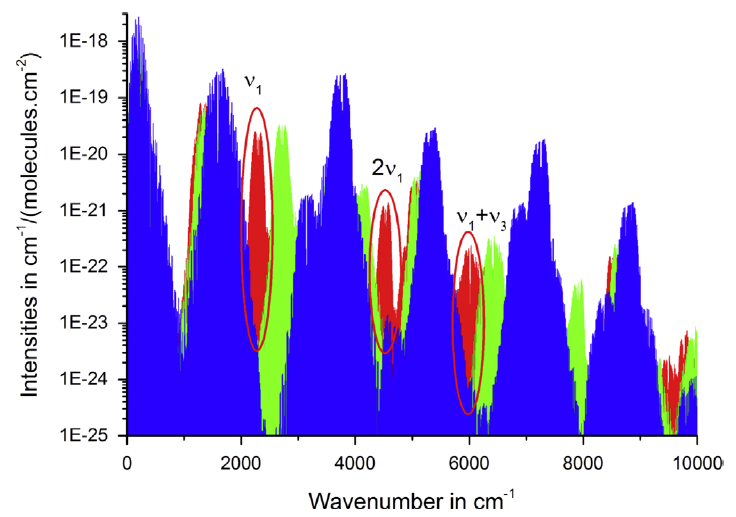}
   \caption{\label{fig:H_molecules_tritium} Rotation–vibration transitions of water isotopologues in the 0 and 10,000 cm$^{-1}$ spectral region; \ce{H_2 ^16O}; \ce{H_2 ^17O} and \ce{H_2 ^18O} (blue), HDO, i.e., \ce{^1H^2HO} (green) and HTO, i.e., \ce{^1H^3HO}, (red). The near IR around 4592~cm$^{-1}$, corresponding to $\sim$2.172~$\mu$m, was used for CRDS detection of HTO with a narrow linewidth distributed feedback diode laser. Figure reproduced from Reference \cite{BrayNuclInst2015}.} 
\end{figure}

Absorption spectroscopy of protium-deuterium isotopes in gas-phase molecular species is relatively common using infrared spectroscopy methods. Figure \ref{fig:H_molecules_tritium} shows calculated cross-sections for rotation-vibration transitions of various gas-phase water isotopologues, highlighting the large molecular isotope shifts and higher cross-sections at infrared wavelengths \cite{BrayNuclInst2015}. Detection of HDO has been demonstrated using optical absorption techniques including  FTIR \cite{griffith2006ftirHDO, LitvakRSCAdv2018, grossleFST2015, YangACS2025}, TD-LAS \cite{mcmanus2015tdlasHDO}, quantum cascade laser absorption spectroscopy (QC-LAS) \cite{BrumfieldPhotonics2016, joly2006dfbqclHDO}, wavelength modulation spectroscopy (WMS) \cite{gianfrani2003wmsHDO}, integrated cavity output spectroscopy (ICOS) \cite{moyer2008icosHDO}, cavity ringdown spectroscopy (CRDS) \cite{gupta2009crdsHDO}, and dual frequency comb spectroscopy \cite{herman2023dcsHDO}. Absorption spectroscopy measurements of deuterium substitutions for protium have also been done for CH$_4$ \cite{campargue2023high}, NH$_3$ \cite{zhu2024nh3, kang2007optical}, and many more. As discussed in Section \ref{isotope_shifts}, molecular isotope shifts result from variations in reduced mass and for H isotopes the relative mass changes are large, leading to corresponding isotope shifts that are much larger than the widths of the rotational lines at atmospheric pressures and temperatures. In many cases, vibrational isotope shifts are sufficiently large to observe shifted bands even in larger polyatomic molecules.  For example, deuterium substitution in gas-phase methanol (CH$_3$O$^{1,2}$H) and ethanol (CH$_3$CH$_2$O$^{1,2}$H) result in clearly distinct and shifted IR absorption bands \cite{brumfieldAnalyst2017}.

While absorption spectroscopy measurements of gas-phase molecules containing deuterium are extensive, reports showing detection of tritium in molecules are more limited. Broadband and high-resolution FTIR studies of gas-phase tritiated water isotopologues have been reported, providing valuable information needed for spectral line parameters.  Hermann \textit{et al.} \cite{hermannJQSRT2021,hermannMolP2022,hermannMolP2023} have reported extensive broadband spectral data on high-resolution FTIR absorption spectra of various tritiated water vapor species. Muller \textit{et al.} \cite{mullerOE2019} developed a custom-built light-pipe cell for high-resolution infrared absorption spectroscopy of HTO (i.e., \ce{^1H^3HO}).  Reinking \textit{et al.} \cite{reinkingJQSRT2019} performed FTIR absorption spectroscopy and analysis of an HTO band near 3.8 $\mu$m. 

As with detection of gas-phase HDO (i.e., \ce{^1H^2HO}), tunable laser-based absorption methods can provide high spectral resolution and high sensitivity for HTO. Bray \textit{et al.} \cite{BrayNuclInst2015} used CRDS with a tunable diode laser near 2.2 $\mu$m wavelengths to detect HTO. Iwamoto \textit{et al.} \cite{iwamotoJJAP2023} also reported HTO detection via CRDS.  Cherrier and Reid \cite{cherrierNucIns1987} performed TD-LAS with a lead salt diode laser and a multipass White cell to measure HTO at infrared wavelengths between 7.2-8.4 $\mu$m .  Karhu \textit{et al.} \cite{karhuPhAc2023} used a photoacoustic spectroscopy method with a QCL near 7.32 $\mu$m  for HTO detection. Cozijn \textit{et al.} \cite{cozijnPRL2024} performed saturated absorption measurements of transitions in the (2-0) band of radioactive tritium hydride using noise-immune cavity-enhanced optical-heterodyne molecular spectroscopy (NICE-OHMS) in the range 1460–1510 nm.

\subsection{Other optical spectroscopy tools}

Several studies have implemented Raman spectroscopy to monitor isotopologues such as \ce{H2}, \ce{^2H2}, \ce{^3H2}, \ce{^1H^2H}, and \ce{^1H^3H} \cite{sturmLP2010, schlosserAC2013, NiemesSensors2021}. Raman spectroscopy has been used to measure isotopic shifts between \ce{^1H_2} and \ce{^2H_2} in dynamic gas streams \cite{FelmyAC2024, KoschnickAC2025}. Similar to other optical diagnostics, Raman spectroscopy offers non-destructive, remote, in situ measurements. Raman spectroscopy of gas-phase tritium-substituted methane has been reported \cite{diazFST2024}, as has Raman spectra of tritiated molecular H \cite{priester2022mura, NiemesSensors2021}.  Because these species have relatively large reduced masses, their Raman shifts are well separated and exhibit little spectral overlap. The primary challenge is achieving sufficient sensitivity given the inherently low Raman scattering cross sections \cite{wenJRS2019}.

\section{Lithium isotopic analysis}
\label{Li_isotopic_analysis}

Optical diagnostics of LPPs have also been applied to the study of Li isotopes using the Li\,I 670.8~nm resonance doublet which possesses the largest isotopic shift of $\approx$15.8 pm among various Li atomic transitions in the UV–VIS spectral region \cite{RadziemskiPRA1995}. Table~\ref{tab:Li_Lit_Review} summarizes
representative studies of Li isotope analysis using optical spectroscopy of LPPs. The studies are organized by diagnostic approach (emission, absorption, and fluorescence) and highlight the diversity of sample matrices, and experimental parameters investigated for Li isotopic measurements. For each entry, the table lists the target material, key experimental
details, and corresponding reference(s).

\begin{table*}
\caption{Summary of literature reporting analysis of Li isotopes (\ce{^6Li} and \ce{^7Li}) via optical (atomic) spectroscopy of LPPs in the UV-VIS spectral region, including information on target composition and experimental conditions, organized by approach (emission versus absorption/fluorescence). All isotopic analysis was performed using the Li I resonance transition centered at $\approx$670.8 nm. Acronyms used in the table are defined as follows: LPP (laser-produced plasma),
UV--VIS (ultraviolet--visible), POE (poly(ethylene oxide)), LiTFSI
(lithium bis(trifluoromethylsulfonyl)imide), Nd:YAG
(neodymium-doped yttrium aluminum garnet), LIBS
(laser-induced breakdown spectroscopy), LIBRIS
(laser-induced breakdown self-reversal isotopic spectrometry), LPV-LIBS (laser-produced-vapor LIBS), TD-LAS (tunable diode laser absorption spectroscopy), LIF (laser-induced fluorescence), OPO (optical parametric oscillator), BB-AS (broadband absorption spectroscopy), SC (supercontinuum), SC-AS (supercontinuum absorption spectroscopy), ML (machine learning), RMSE (root-mean-square error), LOD (limit of detection), CF (calibration-free), and NIST (National Institute of Standards and Technology).}
\label{tab:Li_Lit_Review}
\centering
\renewcommand{\arraystretch}{1.18}
\setlength{\tabcolsep}{4pt}

\begin{tabular}{p{1.0cm} p{2.2cm} p{8.4cm} p{1.5cm}} 
\hline
 & \textbf{Target} &
   \textbf{Experiment details / remarks} &
   \textbf{Ref.} \\
\hline
\multirow{14}{*}{\rotatebox[origin=c]{90}{\textbf{Emission}}}

& \multirow{3}{*}{\ce{LiAlO2}}
& Ar, He, \ce{N2} (0.1--100 Torr), Nd:YAG
(1064 nm, $\approx$6 ns, $\sim$12 J/cm$^2$); lower pressure reduces linewidth and self-reversal
& \cite{KautzOE2023, KarimSCAB2025, HarilalSPIE2024} \\

& 
& He (0.7 Torr), Nd:YAG (1064 nm, $\approx$6 ns, 2.55 J/cm$^2$); Li fine-structure and isotope components partially resolved
& \cite{caleb2026} \\
\cline{2-4}

& \makecell[l]{\ce{Li2TiO3} \\ \ce{Li2CO3}}
& Air (75--760 Torr), LIBS, Nd:YAG (532 nm); Li doublet/isotopic features resolved at lower pressures
& \cite{LaiJACS2026} \\

& \ce{Li2CO3}
& Air/Ar (atmospheric pressure), Nd:YAG
(266/1064 nm, 4 ns, $\approx$190 J/cm$^2$), LIBRIS; self-reversal dip enables \ce{^6Li} quantification with $\approx$6\% relative uncertainty
& \cite{TouchetSCAB2020} \\

& \makecell[l]{\ce{Li2CO3} \\ POE/LiTFSI}
& Air/Ar (atmospheric pressure), Nd:YAG
(266 nm, 4--5 ns). LIBRIS/LIBS; spatially resolved isotopic analysis with 0.74 $\mu$m depth and 3.3 $\mu$m lateral resolution
& \cite{GallotSCAB2023, Gallot2024depth} \\

& LiCl, \ce{Li2CO3}
& Air (atmospheric pressure), Nd:YAG laser
(1064 nm, 7--9 ns, 14 mJ); Li peak shift is noted with enrichment without resolving any structures
& \cite{CremersAS2012} \\

& LiCl
& Air (atmospheric pressure), Nd:YAG
(532/1064 nm, $\approx$10 ns), LPV-LIBS; isotopic shift measured, deep learning corrects self-absorption
& \cite{tran2025isotopic, ShimResultsPhys2025} \\

& \ce{LiOH.H2O}
& \ce{N2} (atmospheric pressure), Nd:YAG
(532 nm, 10 ns, 7.2 J/cm$^2$), LIBRIS; ML-enhanced LIBRIS quantified \ce{^6Li} with RMSE of 5.66 at.\%
& \cite{MoranAO2025} \\

& \ce{LiOH.H2O}
& Ar (0.040 Torr), Nd:YAG
(532 nm, 6 ns, 140 mJ); chemometrics enabled quantitative \ce{^6Li} analysis
& \cite{WoodAppSpec2021} \\

\hline

\multirow{14}{*}{\rotatebox[origin=c]{90}{\textbf{Fluorescence, Absorption}}}
& \ce{Li2C2O4}
& Ar (0.01--1 Torr), excimer ablation laser
(308 nm, $\approx5\times10^7$ W/cm$^2$), tunable dye laser (670.8 nm), LIF; \ce{^6Li}/\ce{^7Li} fine-structure components resolved
& \cite{SmithSCAB1998} \\

& \ce{Li2B4O7}
& Air (atmospheric pressure), Nd:YAG
(1064 nm), diode laser (671 nm), LIBS/TD-LAS; late-time TD-LAS resolves Li fine-structure/isotopic features
& \cite{HullSCAB2021} \\

& Li battery
& He (1 Torr), Nd:YAG (1064 nm), OPO pseudocontinuum  ($\approx$4 ns), BB-AS; Li isotopic/fine-structure components resolved by high-resolution BB-AS
& \cite{MertenSCAB2018} \\

& NIST glass
& Air (5 Torr), Nd:YAG (1064 nm, $\approx$6 ns, 36 mJ), diode laser (671 nm), TD-LAS; Li LOD $\sim$ 6 ppb
& \cite{PhillipsOL2025} \\

& \ce{LiAlO2}
& Air (1 Torr), Nd:YAG
(1064 nm, $\approx$6 ns, 18 mJ), diode laser (671 nm), TD-LAS; Li isotopic/fine-structure components resolved, CF \ce{^6Li}/\ce{^7Li} analysis with 0.6--1.8\% precision
& \cite{PhillipsLiISO2026} \\

& \ce{LiAlO2}
& He (0.7 Torr), Nd:YAG (1064 nm, $\approx$6 ns, 2.55 J/cm$^2$), supercontinuum (SC) light source, SC-AS; instrumental broadening  affects SC-AS measurements
& \cite{caleb2026} \\

\hline

\end{tabular}
\end{table*}


As seen in Table~\ref{tab:Li_Lit_Review}, many of the reported studies of Li isotopic analysis using LPPs rely on optical
emission spectroscopy (LIBS), typically performed under atmospheric or reduced pressure conditions. LIBS offers experimental simplicity and is well suited for direct analysis of solid materials; however, several line-broadening mechanisms and optical thickness effects, including self-absorption and self-reversal (see Section~\ref{line_broadening}), can influence measured emission profiles. Consequently, much of the literature on Li isotopic analysis using LIBS has focused on extracting
isotopic information from partially resolved spectral features of the Li\,I 670.8~nm transition. In the following sections, we first discuss emission-based approaches before turning to absorption and fluorescence techniques that can mitigate or overcome several intrinsic limitations of emission spectroscopy.

\subsection{Laser Induced Breakdown Spectroscopy}
\label{Li_LIBS}

LIBS has been used for both Li elemental and isotopic detection. Reported limits of detection for elemental Li typically range from 0.2--40~ppm with single-pulse LIBS \cite{khater2013trace, audren2026towards}, while double-pulse configurations in aqueous solutions have demonstrated sub-ppm sensitivity \cite{LeeAC2011}. However, it is important to distinguish elemental detection from isotopic discrimination. Although LIBS can detect Li at trace concentrations, resolving the finely spaced isotopic shift between \ce{^6Li} and \ce{^7Li} presents a substantially greater challenge.

The Li I 670.8nm resonance transition (2s–2p) is primarily used for Li isotopic analysis because it exhibits the largest isotopic shift among Li transitions in the UV–VIS region. Achieving sufficient thermal excitation requires elevated plasma temperatures, which in turn increase Doppler broadening. At plasma temperatures near 4000 K—typical of the early evolution of a LIBS plasma—the Doppler width of the Li I 670.8 nm line approaches $\sim$ 10 pm. This magnitude is comparable to the \ce{^6Li}/\ce{^7Li} isotopic separation and is further complicated by overlap with the D1 and D2 fine-structure components. When instrumental broadening from the spectroscopic system is also included, the effective linewidth can approach or exceed the isotopic shift. Consequently, under atmospheric‑pressure conditions, direct spectral resolution of the \ce{^6Li} and \ce{^7Li} components is generally not feasible, even when high‑resolution echelle spectrographs are employed \cite{CremersAS2012}. 

\begin{figure}
    \centering
    \includegraphics[width=1\linewidth]{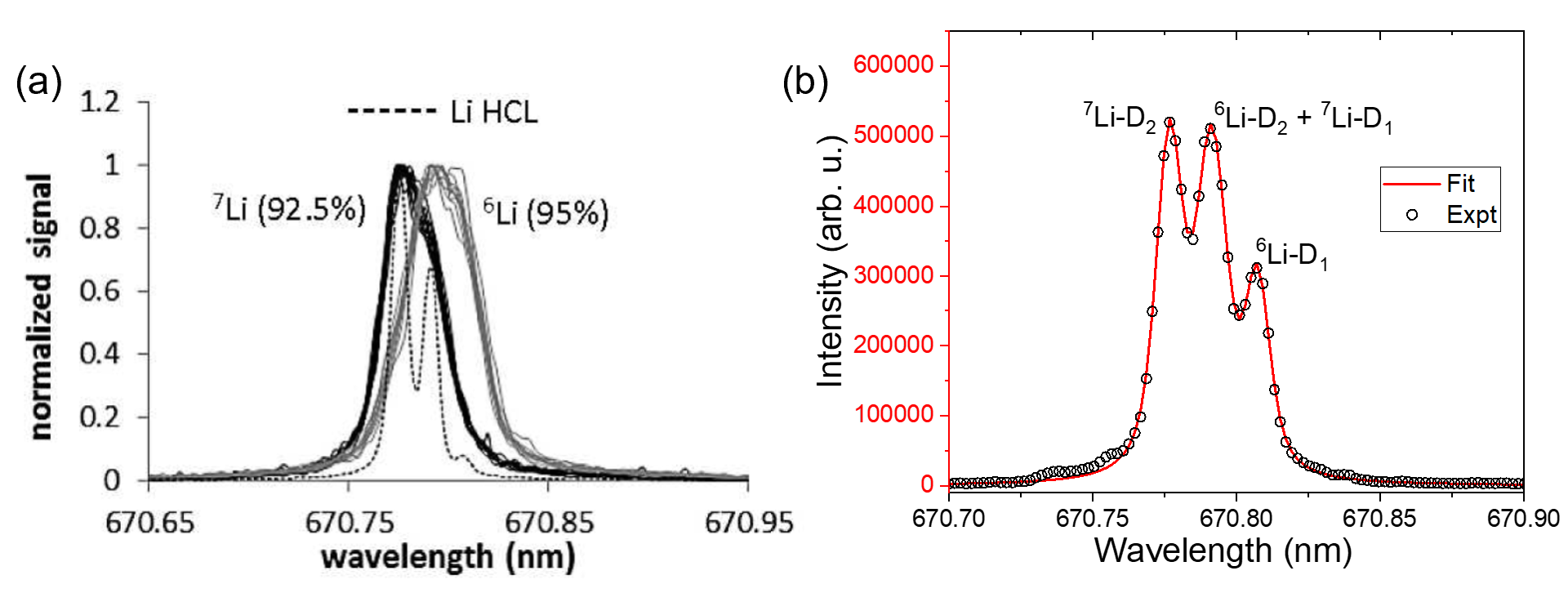}
\caption{\label{fig:Li_isotopes_emission} Emission spectra collected at varying spectral resolutions for the Li\,I 670.8~nm doublet. (a) LIBS spectra obtained at atmospheric pressure from LiCl enriched in \ce{^7Li} (92.5~\% \ce{^7Li}) and \ce{Li2CO3} enriched in \ce{^6Li} (95~\% \ce{^6Li}), shown with a hollow-cathode lamp reference spectrum (HCL, 92.5~\% \ce{^7Li}) \cite{CremersAS2012}. The LIBS measurements were performed using a 1064~nm diode pumped 
solid state laser operated at 14~mJ/pulse, 10~Hz repetition rate, and 
7--9~ns pulse width, with a high-resolution spectrograph with resolving 
power \(\lambda/\Delta\lambda \approx 7.5\times10^{4}\). (b) Emission spectra from LPPs generated from a \ce{LiAlO2} target with a \ce{^6Li}/\ce{^7Li} ratio of $\approx$0.354 (measured by ICP--MS). LPPs were generated using a 1064~nm, $\approx$6~ns Nd:YAG laser at 5~mJ and 10~Hz, corresponding to a fluence of \(\approx 2.55~\mathrm{J/cm^2}\). Spectra were collected in flowing He at 0.7~Torr, 20~mm from the target surface, using a gate delay/width of 50~$\mu$s/5~$\mu$s, and averaged over 20 laser shots. A double-Echelle spectrograph with $\approx$7.2~pm instrumental resolution at 671~nm, corresponding to \(\lambda/\Delta\lambda \approx 9.3\times10^{4}\), was used \cite{caleb2026}.}
\end{figure}

Despite these limitations, measurable changes in the Li\,I 670.8~nm transition have been observed for samples with large differences in isotopic composition \cite{CremersAS2012, WoodAppSpec2021, KautzOE2023, LaiJACS2026}.  In atmospheric pressure air isotopically enriched samples exhibit a measurable shift of the Li\,I 670.8~nm feature; however, the D$_1$/D$_2$ fine-structure components are not spectrally resolved (see Figure~\ref{fig:Li_isotopes_emission}(a)). One practical strategy for improving Li isotopic analysis by LIBS is to reduce Doppler broadening by generating plasmas under reduced-pressure conditions. Under these conditions, rapid plasma expansion promotes cooling and reduction in electron density. Figure~12(b) compares the Li I LIBS spectrum acquired under reduced-pressure conditions (0.7~Torr He) \cite{caleb2026} using a spectroscopic detection systems with similar resolving power ($\approx 8\times10^{4}$)  used in Figure~12(a). 
The Li spectral feature recorded at reduced pressure conditions shows partially-resolved fine structure components with isotopic shifts.

LIBS of Li 670.8 nm  exhibits self-absorption and/or self-reversal because of being a resonance transition with a large oscillator strength. These effects are especially pronounced in confined plasmas at atmospheric pressure and at elevated analyte concentrations. LPPs are inherently inhomogeneous and exhibit spatial and temporal gradients in temperature and electron density \cite{HarilalRMP2022}. Radiative transfer through cooler outer plasma layers can induce self-reversal in strong emission lines, appearing as a central dip in the emission profile \cite{palleschi2022avoiding, Hahn2010, KautzOE2023, KarimSCAB2025}. Reduced pressure can also decrease, although not necessarily eliminate, self-absorption and self-reversal. Motivated by these advantages, several studies have investigated Li in LPPs under reduced-pressure conditions \cite{LaiJACS2026, WoodAppSpec2021, KautzOE2023}, as reflected by the range of background gases and pressures summarized in Table~\ref{tab:Li_Lit_Review}. For example, prior work combined reduced He pressure with chemometric analysis and moderate spectral resolution ($\approx 5.6\times10^{4}$) to predict Li isotope ratios with a mean absolute percent error of approximately 4~\% \cite{KautzOE2023}. 

\begin{figure}
    \centering
    \includegraphics[width=1\linewidth]{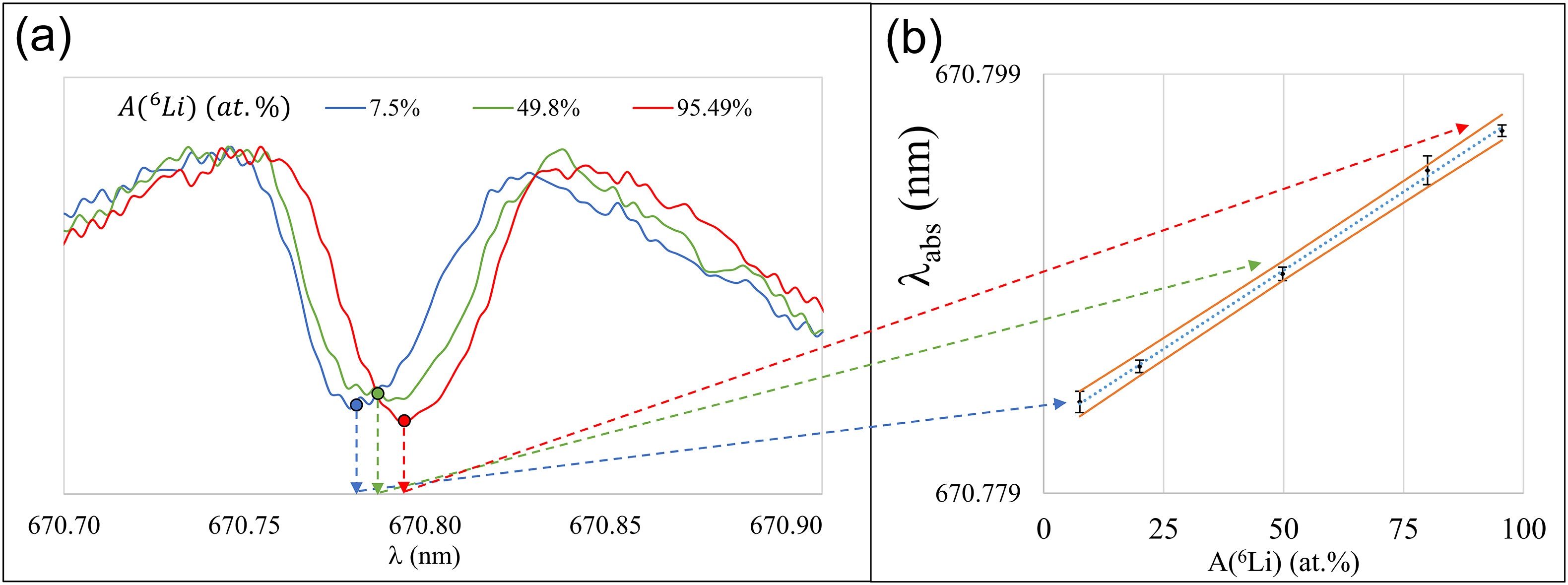}
    \caption{\label{fig:Li_LIBRIS} Illustration of the LIBRIS principle. (a) LIBS spectra acquired from \ce{Li2CO3} samples spanning a range of \ce{^6Li} abundances, showing the systematic red shift of the Li\,I 670.8~nm self-reversal dip with increasing \ce{^6Li} fraction. Measurements were performed at atmospheric pressure using 4~ns laser pulses with an energy of $\approx$5~mJ, and a spectrograph-ICCD with  resolution of \(12.5 \pm 0.5\)~pm (\(\lambda/\Delta\lambda \approx 5.4\times10^{4}\)) at 670.776~nm. (b) Calibration curve relating the measured self-reversal dip wavelength (\(\lambda_{\text{abs}}\)) to the measured atomic \ce{^6Li} abundance \cite{TouchetSCAB2020}.}
\end{figure}

The self‑reversal observed in the Li I 670.8 nm emission line under atmospheric conditions can be leveraged for isotopic analysis. Touchet \textit{et al.} \cite{TouchetSCAB2020} demonstrated that the wavelength of the absorption dip in a self‑reversed profile is linearly correlated with Li isotopic composition, introducing the method known as laser‑induced breakdown self‑reversal isotopic spectrometry (LIBRIS).  As shown in Figure~\ref{fig:Li_LIBRIS}(a), the position of the self-reversal dip shifts systematically with increasing \ce{^6Li} fraction. The corresponding calibration curve in Figure~\ref{fig:Li_LIBRIS}(b) demonstrates a linear correlation between the central dip wavelength, $\lambda_{\text{abs}}$, and the \ce{^6Li} concentration, enabling quantitative isotopic analysis. Subsequent studies have extended LIBRIS to spatial mapping \cite{GallotSCAB2023}, depth profiling \cite{Gallot2024depth}, and machine-learning-enhanced analysis \cite{MoranAO2025}, with reported relative uncertainties ranging from 2.45~\% to 40~\%, depending on sample composition and plasma conditions.

It is also worth noting that isotopic shifts of Li-containing diatomic molecules such as LiO and LiH have been reported in molecular spectroscopy studies \cite{stwalley1993spectroscopy, nocunJMS2001}. However, to date, LPPs have not been systematically utilized as controlled molecular reservoirs for Li isotopic analysis. While molecular emission pathways could offer alternative isotopic signatures, their formation, stability, and detectability in transient LPPs may be more challenging than studying Li atomic transitions in the UV–VIS spectral region.

While emission-based methods are effective for rapid Li detection and screening, accurate isotopic analysis remains fundamentally constrained by three key factors: (1) Doppler broadening, which is significant for light elements such as Li; (2) instrumental resolution limits of light-detection systems; and (3) optical-thickness effects that lead to self-absorption and self-reversal. Techniques such as LIBRIS creatively exploit self-reversal as an isotopic discriminator; however, the position and depth of the reversal minimum remain coupled to plasma temperature gradients and optical thickness. These intrinsic limitations of emission spectroscopy have motivated increasing interest in absorption- and fluorescence-based diagnostics, which probe lower-energy-state populations.

\subsection{Laser Induced Fluorescence}
\label{Li_LIF}

\begin{figure}
    \centering
    \includegraphics[width=0.75\linewidth]{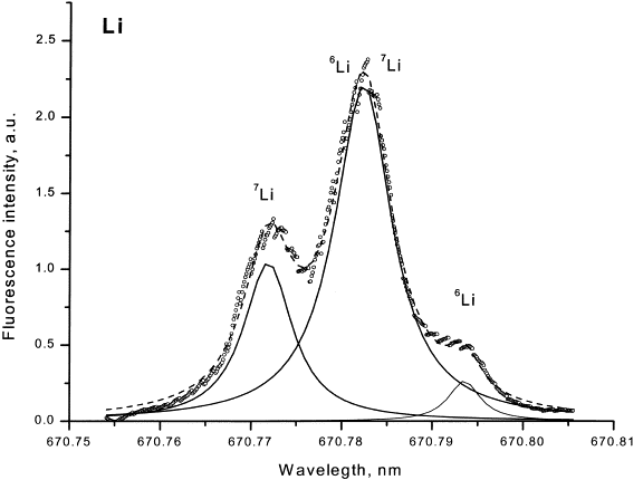}
    \caption{\label{fig:Li_LIF} LIF spectrum of the Li\,I 670.8~nm doublet obtained from a LPP generated from a pressed lithium oxalate pellet. The plasma was produced using a 308~nm excimer laser at an irradiance of \(\approx 5\times10^{7}\)~W\,cm\(^{-2}\). A dye laser with a linewidth of \(\approx 10\)~pm was scanned across the Li\,I 670.8~nm transition at 25~pm\,min\(^{-1}\), with the excitation beam passing through the plasma $\approx$1~cm above the sample surface. Resonance fluorescence was collected perpendicular to the excitation beam and detected using a monochromator--PMT. Dots represent the experimental data, solid lines show the deconvolved Lorentzian components, and the dashed line represents the multipeak Lorentzian fit. Figure reproduced from Reference ~\cite{SmithSCAB1998}.} 
\end{figure}

Absorption-based techniques have been increasingly explored for Li isotopic analysis, as summarized in  Table~\ref{tab:Li_Lit_Review}. By probing ground or lower state populations at delayed times, absorption-based measurements can probe the LPP at lower temperatures, and therefore narrower Doppler linewidths. LIF has been explored to a limited extent for Li isotopic analysis in LPPs. Smith \textit{et al.} \cite{SmithSCAB1998} reported LIF measurements of Li in LPPs generated from \ce{Li2C2O4} targets in reduced-pressure Ar gas (mTorr level), using an excimer-pumped dye laser tuned to the Li\,I 670.8~nm resonance transition. Excitation spectra (an example of which is given in Figure \ref{fig:Li_LIF}) shows clearly resolved D$_1$ and D$_2$ fine-structure components \ce{^6Li} and \ce{^7Li}, demonstrating promise for isotopic detection via LIF of LPPs. Despite this early demonstration, systematic studies of Li isotope ratio quantification via LIF of LPPs have not been reported, and the technique remains significantly less developed than LIBS or absorption-based approaches for Li isotopic analysis.


\subsection{Absorption Spectroscopy} 
\label{Li_AS}

Compared to LIF, absorption spectroscopy techniques have been investigated more extensively for Li isotopic analysis, as summarized in  Table~\ref{tab:Li_Lit_Review}. This table includes BB-AS, and  TD-LAS techniques applied to LPPs. Furthermore, absorption methods provide a quantitative measurement of atomic column density and are not subject to self-absorption effects that complicate interpretation of the Li\,I 670.8~nm resonance doublet in LIBS. 

In absorption spectroscopy, the measured area of a spectral line is directly proportional to the column density in the lower state of the probed transition \cite{HarilalRMP2022}.  For Li, the large energy separation between ground and excited states results in most population existing in the ground state as the LPP cools. To determine the associated ground state column densities of \ce{^6Li} and \ce{^7Li}, and their ratio, peak fitting methods can be used to determine the areas of the spectral lines. Models for the spectrum can be constructed by including the fine and hyperfine structure \cite{PhillipsLiISO2026} or by approximating the lines as arising from single transitions \cite{PhillipsOL2025, caleb2026, MertenSCAB2018}, with the choice likely depending on the level of accuracy and precision required. 

Both BB-AS and TD-LAS have been applied to Li isotope measurements in conjunction with LPPs \cite{MertenSCAB2018, HullSCAB2021,PhillipsLiISO2026,caleb2026}. In BB-AS, spectral resolution is limited by the light detection system, and isotopic analysis requires multi-peak fitting of the partially resolved D$_1$/D$_2$ fine-structure components and their associated isotopic shifts. The measured lineshape represents a convolution of intrinsic broadening mechanisms and instrumental response, making accurate spectral modeling essential for extracting \ce{^6Li}/\ce{^7Li} ratios.

\begin{figure}
    \centering
    \includegraphics[width=1\linewidth]{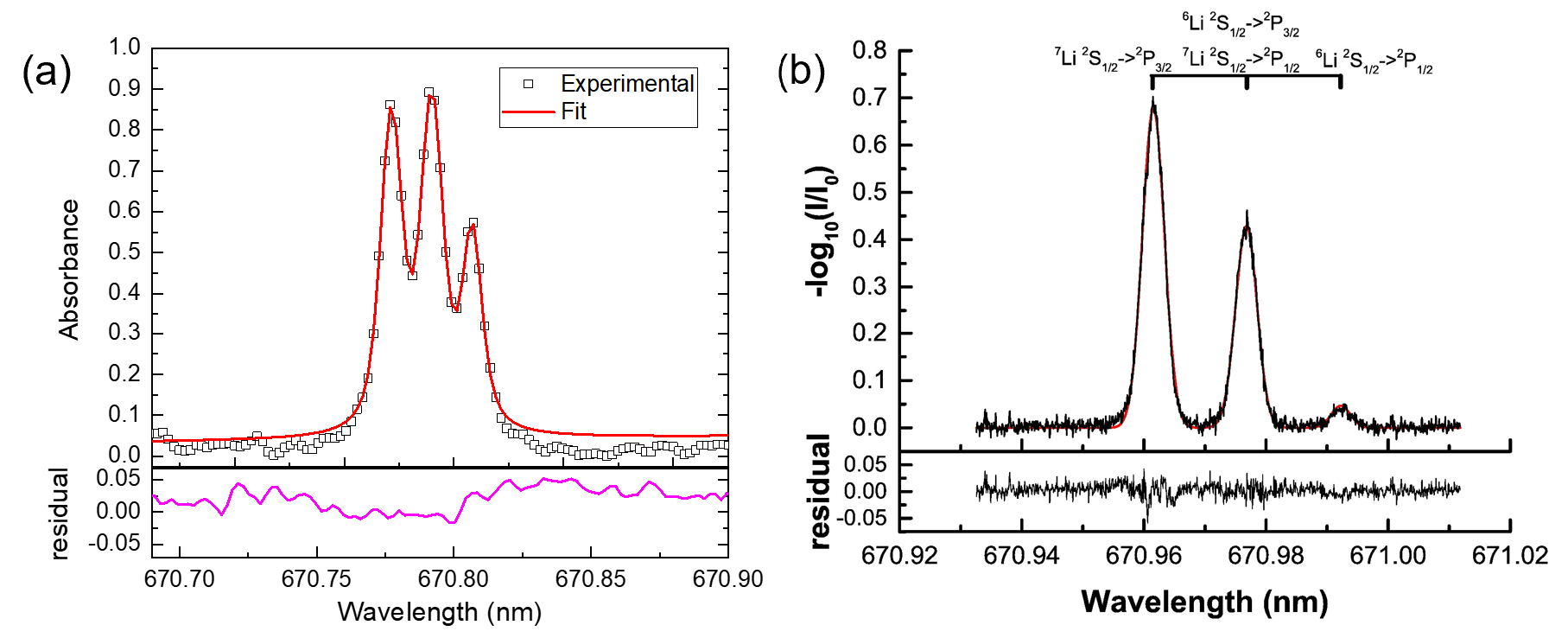}
    \caption{\label{fig:Li_absorption} Absorption spectra of the Li\,I 670.8~nm doublet acquired with light-detection systems of differing resolving power. (a) BBAS of LPPs generated in flowing 0.7~Torr He. The output from a supercontinuum source (NKT Photonics SuperK Compact) was passed through the plasma 20~mm above a \ce{LiAlO2} target with a \ce{^6Li}/\ce{^7Li} ratio of 0.354. Spectra were recorded with a gate delay/width of 50~\(\mu\)s / 5~\(\mu\)s and averaged over 20 laser shots using an echelle spectrograph with resolving power \(\lambda/\Delta\lambda \approx 9.0\times10^{4}\). (b) Absorption spectrum of Li\,I 670.8~nm recorded using a pseudo-continuum OPO source in conjunction with an LPP generated in 1~Torr He. The OPO provided \(\approx 4\)~ns pulses with a bandwidth of \(\approx 3~\mathrm{cm^{-1}}\), which were coupled into a photonic-crystal fiber to generate the pseudo-continuum probe. Probe and reference spectra were recorded simultaneously using a double-pass echelle spectrograph--ICCD system with a resolving power of \(\lambda/\Delta\lambda \approx 7.6\times10^{5}\). The spectrum, obtained from a sample of unknown purity and isotopic composition, was fitted using a multi-peak model \cite{MertenSCAB2018}. For both panels, residuals from the spectral fits are shown below the main traces.}
\end{figure}

Representative BB absorption spectra acquired with light detection systems with different resolving powers are shown in Figure~\ref{fig:Li_absorption}. In Figure~\ref{fig:Li_absorption}(a), a supercontinuum laser source combined with an echelle spectrograph (\(\lambda/\Delta\lambda \approx 9\times10^{4}\)) enabled partial separation of the three peaks (\ce{^{7}Li}~D$_2$, \ce{^{6}Li}~D$_2$ + \ce{^{7}Li}~D$_1$, and \ce{^{6}Li}~D$_1$) when LPPs were generated in reduced pressure He gas. In contrast, Figure~\ref{fig:Li_absorption}(b) reports a spectra with even better resolution obtained using an OPO-based pseudo-continuum source and a spectrograph with \(\lambda/\Delta\lambda \approx 7.6\times10^{5}\) \cite{MertenSCAB2018}. The residuals shown beneath each spectrum highlight how well the applied multi-peak fitting compares to experimental data. Quantities such as peak areas that can be extracted from these fits can subsequently be used for quantitative isotopic analysis.

\begin{figure}
    \centering
    \includegraphics[width=1\linewidth]{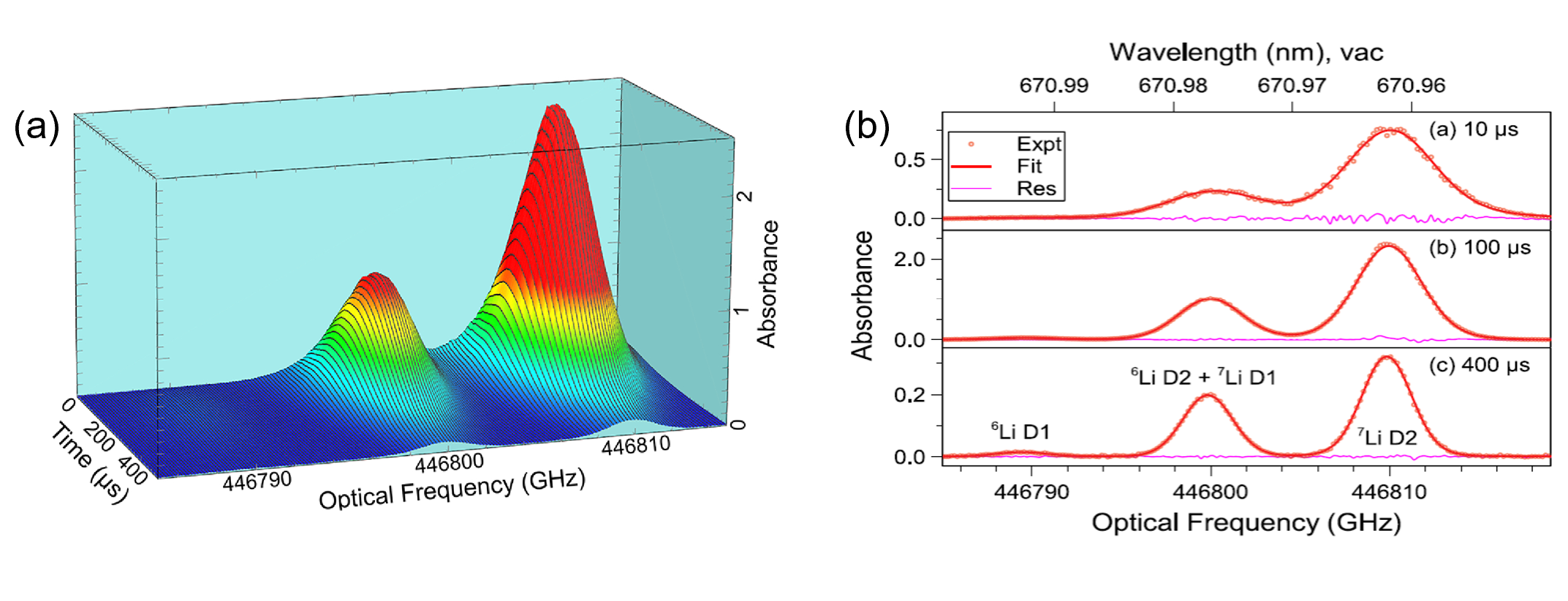}
    \caption{\label{fig:Li_TDLAS} TD-LAS measurements of the Li\,I 670.8~nm D\(_1\) and D\(_2\) transitions following laser ablation of NIST SRM~610. LPPs were generated in 5~Torr air using a 1064~nm, \(\approx 6\)~ns Nd:YAG laser at \(\approx 36\)~mJ and 10~Hz. An external-cavity diode laser probed the LPP \(\approx 4\)~mm above the sample and was scanned across the Li\,I transitions in 0.3~pm (180~MHz) steps. (a) Time-resolved absorption map showing the evolution of the Li\,I doublet over 0–450~\(\mu\)s. (b) Representative absorbance spectra extracted at delay times of 10, 100, and 400~\(\mu\)s. Open circles denote experimental data, solid red curves represent multi-peak spectral fits, and magenta traces correspond to fit residuals. Features associated with \ce{^6Li} and \ce{^7Li} D\(_1\) and D\(_2\) transitions are indicated. Reprinted with permission from \cite{PhillipsOL2025} \textcopyright Optica Publishing Group.}
\end{figure}

In contrast to BB-AS, TD-LAS employs a narrow-linewidth probe laser scanned across the Li\,I D$_1$/D$_2$ transitions, providing substantially higher spectral resolution and improved precision in isotopic discrimination. For example, a laser with 10 MHz linewidth provides a resolving power of $>10^{7}$ at 671 nm, which is much smaller than the linewidths of the transitions. Hull \textit{et al.} \cite{HullSCAB2021} measured TD-LAS of Li under atmospheric pressure air, and showed that spectral linewidths at late time in LPP evolution were small enough to resolve the fine structure and isotope splitting of the Li D1 and D2 transitions. TD-LAS experiments are often performed at reduced pressures to reduce noise and extend the duration of the absorption signals.  Time-resolved TD-LAS measurements of Li in a LPP under 5 Torr air are presented in Figure~\ref{fig:Li_TDLAS}. The absorption map in Figure~\ref{fig:Li_TDLAS}(a) illustrates the temporal evolution of the Li\,I spectrum following ablation, while Figure~\ref{fig:Li_TDLAS}(b) shows representative absorbance spectra extracted at delay times of 10, 100, and 400~\(\mu\)s. As the plasma cools and expands, linewidths decrease and isotopic features become increasingly distinct. Phillips \textit{et al.} \cite{PhillipsOL2025} investigated signal averaging considerations and reported an elemental limit of detection of 180~ppb in single-shot operation, improving to 6~ppb with averaging over 1000 ablation events.

\begin{figure}
    \centering
\includegraphics[width=1\linewidth]{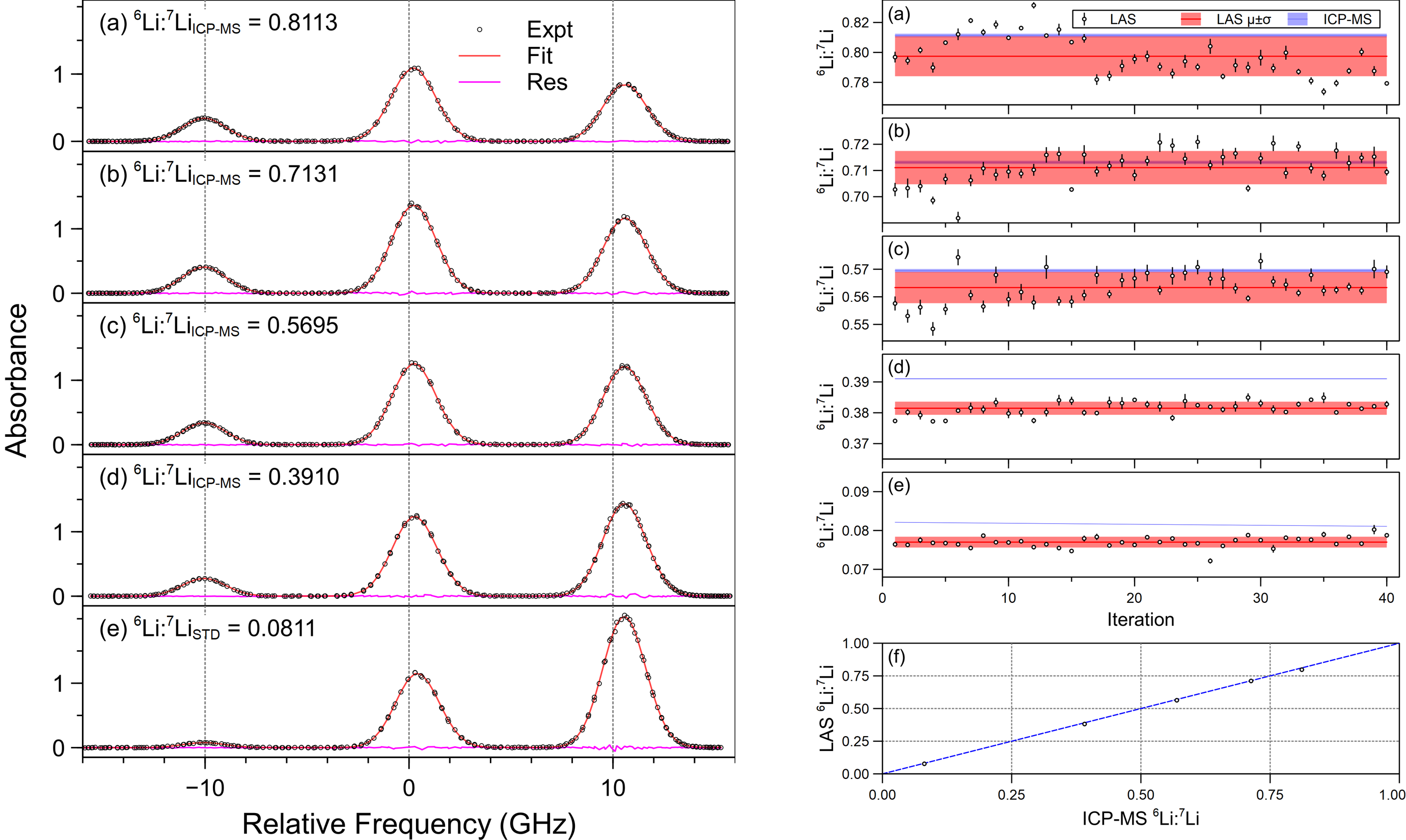}
    \caption{\label{fig:Li_LAS_ISO} Performance of TD-LAS for Li isotope ratio measurements. LPPs were generated from \ce{LiAlO2} samples in 1~Torr air using a 1064~nm, \(\approx 6\)~ns Nd:YAG laser at \(\approx 18\)~mJ and 10~Hz. A narrow-linewidth (\(<10\)~MHz) ECDL probed the LPP \(\approx 5\)~mm above the sample surface. (left) Measured and fit absorption spectra using LAS on samples with varying \ce{^6Li}:\ce{^7Li} isotope ratios, with 30 s acquisition times. (right) LAS calibration-free isotope ratios determined from the spectral fits show good agreement with ratios measured by ICP-MS, and relative isotopic precisions $\approx$ 1\%.  Figure adapted from Reference \cite{PhillipsLiISO2026}.}
\end{figure}

Overall, absorption spectroscopy provides improved spectral discrimination relative to emission-based methods by   enabling delayed-time probing of Li atoms at lower temperatures and narrower linewidths, and by eliminating instrumental broadening in the case of TD-LAS. Quantification abilities of LAS are also expected to be higher than LIBS. Recent efforts are investigating analytical performance of TD-LAS for determining isotope ratios of Li without requiring calibration to external standards. Figure~\ref{fig:Li_LAS_ISO} shows results from Phillips \textit{et al.} \cite{PhillipsLiISO2026} in which TD-LAS was used to measure Li absorption spectra from samples with varying \ce{^6Li}:\ce{^7Li} isotope ratios, each obtained in 30s acquisition times. A physics-based spectral model was used to perform quantitative spectral fitting, and statistical analysis of the isotope ratios determined by TD-LAS showed good agreement with values determined from ICP-MS, especially considering the calibration-free approach of LAS. Relative isotopic precisions $\approx$ 1 \% were demonstrated by LAS, showing the ability of LAS methods to achieve high analytical performance for rapid measurements without sample preparation, fulfilling the primary goals of these optical approaches.

\section{Summary and Outlook}
\label{summary_outlook}

Optical spectroscopy of LPPs, including optical emission, laser absorption, and laser induced fluorescence methods, has emerged as a flexible and increasingly capable platform for isotopic analysis of H and Li from solid target materials. These approaches have progressed substantially from early qualitative observations of isotope splitting toward quantitative methods that, under optimized conditions, can approach or rival the performance of traditional mass spectrometry tools. The breadth of relevant applications spans nuclear energy, defense, safeguards and forensics, geology, and environmental monitoring, all fields where isotopic signatures carry critical information. The following summarizes the current state of the art in H and Li isotopic analysis via optical spectroscopy of LPPs, the benefits and limitations of these methods, and identifies research directions that could substantially extend their analytical capabilities and application space.

\subsection{State-of-the-art in hydrogen and lithium isotopic analysis}

For H, most work involving LPP-based detection has focused on isotopic analysis of the Balmer-$\alpha$ line, but some demonstrations have shown the potential of using \ce{O^1H}/\ce{O^2H} molecular bands \cite{RanSCAB2021}. Emission-based studies demonstrate that clear resolution of \ce{^1H_\alpha}/\ce{^2H_\alpha} \cite{KurniawanAC2006, KurniawanAS2014, KautzOE2021}, and more recently \ce{^1H_\alpha}/\ce{^3H_\alpha} \cite{HarilalJAAS2024}, requires careful management of Stark and Doppler broadening and self-absorption, typically achieved through plasma generation in reduced-pressure environments, delayed gated detection, and, in some cases, microwave \cite{VujadinovicEPhys2026} or double-pulse excitation \cite{KautzSCAB2024, BurgerPoP2018}. These same strategies have been adapted to in situ configurations in fusion devices, where LIBS now supports assessment of fuel retention and impurity deposition on PFCs \cite{SuchovnovaNME2017, VanDerMeidenNuclearFusion2021}. Other methods have shown promising results for gas-phase or plasma-phase detection of H isotopes, including TALIF \cite{schmidt2016femtosecond, KajitaNME2025}, Raman \cite{FelmyAC2024}, and LAS \cite{mullerOE2019}. Cavity-enhanced laser spectroscopy methods have shown detection of \ce{^3H^1HO} with high sensitivity and species selectivity \cite{iwamotoJJAP2023}.

For Li, the comparatively small isotopic shift of the resonance Li I 670.8~nm line imposes stricter demands on spectral resolution and plasma control than are required for H. LIBS performed at moderate to atmospheric pressures cannot resolve the fine-structure components of the Li I doublet (centered at $\approx$670.8~nm), though in some cases integrated peak areas can be used to estimate \ce{^6Li}/\ce{^7Li} ratios when coupled with chemometric analysis \cite{WoodAppSpec2021, KautzOE2023}. Reduced-pressure LIBS leverages the narrower linewidths achievable at lower pressures to improve isotopic discrimination. LIBRIS instead exploits self-reversal of the Li I 670.8~nm line, using the position of the self-reversal dip as a calibration metric for isotopic composition \cite{TouchetSCAB2020}. Absorption-based methods, including BB-AS \cite{caleb2026, MertenSCAB2018} and TD-LAS \cite{HullSCAB2021, PhillipsLiISO2026}, have emerged as particularly powerful for Li isotopic analysis. By probing the Li I resonance transition from the ground state at later times of LPP evolution, when temperatures are lower and Doppler and Stark broadening are reduced, absorption spectroscopy techniques directly access ground-state populations and achieve improved isotopic discrimination \cite{PhillipsLiISO2026}. High-resolution spectrographs and narrow-linewidth diode lasers have allowed partial to full resolution of Li I fine-structure and isotopic splitting, with reported detection limits reaching the low-ppb range under optimized conditions using TD-LAS \cite{PhillipsOL2025}. Physics-based spectral models have been used for quantitative fitting of measured high-resolution absorption spectra to extract isotope ratios without requiring isotopic calibration standards \cite{PhillipsLiISO2026}.

\subsection{Benefits and limitations of current approaches}

Emission-, absorption-, and fluorescence-based approaches have complementary strengths. Emission-based techniques such as LIBS offer broad applicability, multi-element detection, and operational simplicity, but constraints come from isotopic resolution by plasma-induced/instrumental line broadening and self-absorption. Absorption-based methods (BB-AS and TD-LAS) benefit from probing low- and ground-state transitions at reduced Doppler temperatures, enabling improved isotopic discrimination under controlled conditions, though at the cost of more complex experimental configurations. Absorption methods typically achieve higher performance for LPPs generated in reduced pressure environments, but operation at ambient pressures is also possible. LIF offers high selectivity through resonant excitation and provides standoff capability with reduced alignment constraints relative to transmission-based absorption methods. Together, these complementary optical diagnostics highlight the importance of coupling carefully controlled LPP generation conditions, enabled by the choice of laser pulse duration (ns-fs), wavelength, double pulsing etc., with high resolution instrumentation and multimodal spectroscopic approaches to achieve quantitative, reproducible isotopic analysis across diverse sample matrices. 

Despite significant progress, several overarching challenges remain. Many optical isotopic measurements reported to date are demonstration-scale studies, with no standardization of
plasma-generation conditions, calibration strategies, or data-analysis workflows, making direct cross-study comparison difficult and complicating quantitative benchmarking against mass spectrometry techniques. The sensitivity and accuracy of most techniques are also strongly dependent on experimental parameters, including ambient pressure, gas composition, laser energy and pulse timing, and light collection geometry,
requiring careful optimization for each sample matrix and application. 

It is also important to recognize that experimental conditions that improve spectral resolution do not necessarily maximize quantitative precision or accuracy, and the factors governing quantitative performance can differ between emission- and absorption-based measurements. In addition, the use of a high resolution spectrograph or narrow linewidth laser may enable discrimination of closely spaced isotopic shifts, but spectral resolution alone does not ensure accurate and precise isotope quantification. Therefore, advancing isotopic analysis using optical spectroscopy of LPPs requires balancing spectral resolution with signal-to-noise ratio and measurement stability, together with a fundamental understanding of line broadening, optical depth, and other processes that influence quantitative performance.

Isotopic analysis of tritium, and of Li isotopes in chemically complex or heterogeneous matrices, remains in early stages; relatively few studies have addressed matrix effects, neutron-irradiated samples, or the long-term stability and reproducibility of these methods, particularly in environments where high temperatures or radiation is present. These gaps represent the primary barriers to transitioning optical LPP diagnostics from laboratory demonstrations to operational, standard analytical tools.

\subsection{Future directions}

Several research directions could advance optical spectroscopy of LPPs for H and Li isotopic detection and analysis. Related to measurement science, systematic studies that map isotopic precision, accuracy, and detection limits as a function of plasma conditions and resolution (i.e., spectral resolution constraints related to the instrumental broadening for broadband methods) are needed for both H and Li. The achievable precision and detection limits should also be evaluated across a broad range of isotope abundance ratios, as measuring extremes of isotope ratios can be challenging in the presence of noise or strong spectral overlap. The extent of this limitation depends on the signal-to-noise ratio, spectral resolution, and the ability of spectral fitting methods to distinguish overlapping peaks.

Improving existing methods or developing new sensing approaches using LAS will help improve precision, accuracy, and sensitivity for isotopic detection in both LPPs and gas phase molecular species. Data measured using optical spectroscopy should be benchmarked against isotopic standards or reference measurements using mass spectrometry methods for the same samples. Inter-laboratory comparison studies using common reference materials with known isotope ratios would also be useful for evaluating measurement reproducibility across different instruments and experimental configurations. Such studies could help establish standardized approaches for reporting plasma-generation and detection conditions, calibration procedures, data-analysis methods, and measurement uncertainty, allowing more direct comparison of quantitative results across laboratories.

Instrumentation development for LIBS should prioritize compact, high-resolution spectrographs, including double-echelle, Virtually Imaged Phased Array (VIPA), and spatial heterodyne designs. For LAS methods, improved laser sources and integrated sensing systems will help reduce system size and complexity, enabling field-deployable measurement configurations. While portable and handheld LIBS instruments are commercially available, their spectral resolution (typically on the order of 0.05--0.1 nm) is generally insufficient to resolve the isotope shifts discussed in this review. In addition, commercially available field-deployable LIF and LAS systems designed specifically for optical isotope analysis of LPPs are not currently available, although TD-LAS instruments for gas sensing of hydrogen isotopes in molecular isotopologues do exist. The development of compact, high-resolution, active spectroscopic systems capable of standoff isotopic sensing with real-time analysis remains an active area of research and development.

For measurements without complete resolution of isotope peaks, continued advances in data analysis offer promising routes to quantitative analysis, including physics-informed spectral fitting models for overlapping and self-reversed line profiles and machine-learning approaches that extract isotopic information from subtle changes in spectral features. Supervised regression approaches, including partial least squares (PLS) regression and neural-network-based methods, should be evaluated for extracting isotope ratios from partially overlapping spectra, while classification methods may enable rapid identification or screening of samples with different isotope compositions. An important research need is to determine how these models perform across changes in sample matrix, plasma conditions, instrumentation, and measurement geometry, and whether models developed using controlled laboratory measurements can be transferred to field-deployable systems. LIF and two-dimensional fluorescence spectroscopy (2D-FS) \cite{PhillipsSciRep2017,HarilalOL2016,HarilalOE2017} may be promising areas of future study for Li isotopic detection and analysis when standoff detection is needed.

The conditions influencing laser ablation significantly affect plasma formation and can thereby influence isotopic resolution \cite{HariAPR2018}. As such, additional studies are required to systematically modify laser parameters and assess their impact on isotopic resolution. The laser produced colliding plasmas is an emerging field where the stagnation region offers a cooler plasma environment that may be advantageous for isotopic analysis \cite{shilpa2026colliding}.

From our perspective, several research directions are important for advancing isotopic measurements of H and Li using optical spectroscopy of LPPs. First, quantitative performance needs to be better established across a broad range of isotope abundance ratios and experimental conditions, including systematic evaluation of matrix effects and their influence on measurement precision, accuracy, and detection limits. Second, direct comparisons, where possible, among emission- and absorption-based measurements could improve understanding of the relative benefits and limitations of each approach and help determine which technique is best suited for a particular measurement or detection problem. Comparatively limited work has been performed using LIF for Li isotopic analysis, or using molecular species containing Li isotopes, and these approaches may be valuable for some measurements. Finally, advances in the fundamental understanding of optical spectroscopy of LPPs should be translated into high-resolution systems suitable for rapid, online, field, or in situ measurements. Such systems could enable rapid screening and process monitoring or facilitate analysis of radioactive or hazardous materials where sample handling should be minimized. Approaches that integrate complementary emission- and absorption-based measurements may also be beneficial. For example, emission spectroscopy could be used for H isotope measurements while absorption spectroscopy could provide Li isotope ratios from the same sample or material system.

The application space for optical spectroscopy of LPPs is also expanding. Advanced nuclear energy systems, both fission and fusion, increasingly demand diagnostics capable of providing near-real-time feedback for process monitoring and measurements of H and Li isotope inventories for safe operation and regulatory compliance. Nuclear safeguards and forensics require rapid, field-deployable analytical methods, a niche that several of the approaches reviewed here are well positioned to fill as instrumentation matures. Further, Li is a critical mineral whose isotopic composition carries important information in contexts ranging from battery materials research and geochemical tracing to nuclear nonproliferation, where field-deployable isotopic screening methods would offer significant practical advantages over laboratory-based methods. Advancements including hybrid strategies \cite{LyuPhysScripta2021, VSApplSpec2021, WuSCAB2017} or complementary measurements with mass spectrometry or other analytical or imaging methods \cite{FerreiraJAAS2024, KarhunenJNM2015} can provide more comprehensive approaches for light isotope measurements across nuclear energy, forensics, defense, materials science, geochemistry, and environmental science disciplines.

\section*{Glossary of technical terms and acronyms}

\begin{tabular}{ll}
2D-FS & Two-dimensional fluorescence spectroscopy \\

ANN & Artificial neural network \\
APT & Atom probe tomography \\

BB-AS & Broadband absorption spectroscopy \\

CCD & Charge-coupled device \\
CF-LIBS & Calibration-free laser-induced breakdown spectroscopy \\
CRDS & Cavity ringdown spectroscopy \\

DP & Double pulse \\
DP-LIBS & Double-pulse laser-induced breakdown spectroscopy \\
DPSS & Diode-pumped solid-state \\

ECDL & External-cavity diode laser \\
ERDA & Elastic recoil detection analysis \\

FTIR & Fourier transform infrared spectroscopy \\
FWHM & Full width at half maximum \\

GD-MS & Glow discharge mass spectrometry \\

ICCD & Intensified charge-coupled device \\
ICOS & Integrated cavity output spectroscopy \\
ICP-MS & Inductively coupled plasma mass spectrometry \\
IICIS & Integral intensity correction internal standard \\

LA & Laser ablation \\
LAMIS & Laser ablation molecular isotopic spectrometry \\
LAS & Laser absorption spectroscopy \\
LG-SIMS & Large-geometry secondary ion mass spectrometry \\
LIBRIS & Laser-induced breakdown self-reversal isotopic spectrometry \\
LIBS & Laser-induced breakdown spectroscopy \\
LIF & Laser-induced fluorescence \\
LOD & Limit of detection \\
LPP & Laser-produced plasma \\
LSC & Liquid scintillation counting \\
LTE & Local thermodynamic equilibrium \\

MC-ICP-MS & Multi-collector inductively coupled plasma mass spectrometry \\
MIP & Microwave-induced plasma \\
ML & Machine learning \\
MSR & Molten salt reactor \\
MW-LIBS & Microwave-assisted laser-induced breakdown spectroscopy \\

NanoSIMS & Nanoscale secondary ion mass spectrometry \\
NIR & Near-infrared \\
NMR & Nuclear magnetic resonance \\
NR & Neutron reflectometry \\
NRA & Nuclear reaction analysis \\

OES & Optical emission spectroscopy \\
OPO & Optical parametric oscillator \\

PCA & Principal component analysis \\
PLS & Partial least squares \\
PMT & Photomultiplier tube \\

RIMS & Resonance ionization mass spectrometry \\

SC & Supercontinuum \\
SC-AS & Supercontinuum absorption spectroscopy \\
SIMS & Secondary ion mass spectrometry \\
SNR & Signal-to-noise ratio \\
SP & Single pulse \\
SP-LIBS & Single-pulse laser-induced breakdown spectroscopy \\

TD-LAS & Tunable-diode laser absorption spectroscopy \\
TD-MS & Thermal desorption mass spectrometry \\
TEA & Transversely excited atmospheric \\
TIMS & Thermal ionization mass spectrometry \\
ToF-SIMS & Time-of-flight secondary ion mass spectrometry \\

UV--VIS & Ultraviolet--visible \\

WMS & Wavelength modulation spectroscopy \\

\end{tabular}


\ack{PNNL is a multi-program national laboratory operated by Battelle for the U.S. Department of Energy under Contract DE-AC05-76RL01830. The content of the information does not necessarily reflect the position or the policy of the federal government, and no official endorsement should be inferred. }

\funding{DOE/NNSA Tritium Modernization Program, Defense Threat Reduction Agency (HDTRA1-20-2-0001), U.S. DOE NNSA Consortium for Nuclear Forensics (DE-NA0004142), University of Arizona Office of Research \& Partnerships, Technology and Research Initiative Fund: Fusion Energy.}


\bibliography{Refs.bib}
\bibliographystyle{unsrtnat}

\end{document}